\documentclass[%
prx,letterpaper,
superscriptaddress,
 amsmath,amssymb,
 aps,nofootinbib,twocolumn
]{revtex4-2}

\usepackage{graphicx}
\graphicspath{{./Main Figures/}{./Extended Figures/}}
\usepackage{dcolumn}
\usepackage{bm}
\usepackage{physics}
\usepackage{sidecap}
\usepackage{float}
\usepackage{scalerel}
\usepackage{xr}
\usepackage{multibib}
\newcites{methods}{References}
\setcitestyle{super}
\makeatletter

\begin{document}

\title{Many-body cavity quantum electrodynamics with driven inhomogeneous emitters}

\author{Mi Lei}
	\thanks{These two authors contributed equally to this work.}
	\affiliation{Kavli Nanoscience Institute and Thomas J. Watson, Sr., Laboratory of Applied Physics, California Institute of Technology, Pasadena, California 91125, USA}
	\affiliation{Institute for Quantum Information and Matter, California Institute of Technology, Pasadena, California 91125, USA}
\author{Rikuto Fukumori}
	\thanks{These two authors contributed equally to this work.}
	\affiliation{Kavli Nanoscience Institute and Thomas J. Watson, Sr., Laboratory of Applied Physics, California Institute of Technology, Pasadena, California 91125, USA}
	\affiliation{Institute for Quantum Information and Matter, California Institute of Technology, Pasadena, California 91125, USA}
\author{Jake Rochman}
	\affiliation{Kavli Nanoscience Institute and Thomas J. Watson, Sr., Laboratory of Applied Physics, California Institute of Technology, Pasadena, California 91125, USA}
	\affiliation{Institute for Quantum Information and Matter, California Institute of Technology, Pasadena, California 91125, USA}
	\author{Bihui Zhu}
	\affiliation{Homer L. Dodge Department of Physics and Astronomy, The University of Oklahoma, Norman, OK 73019, USA}
    \author{Manuel Endres}
	\affiliation{Institute for Quantum Information and Matter, California Institute of Technology, Pasadena, California 91125, USA}
	\affiliation{Division of Physics, Mathematics and Astronomy, California Institute of Technology, Pasadena, CA 91125, USA}
  \author{Joonhee Choi}
        \thanks{Present address: E. L. Ginzton Laboratory and Department of Electrical Engineering, Stanford University, Stanford, California 94305, USA}
        \email[]{joonhee.choi@stanford.edu}
	\affiliation{Institute for Quantum Information and Matter, California Institute of Technology, Pasadena, California 91125, USA}
	\affiliation{Division of Physics, Mathematics and Astronomy, California Institute of Technology, Pasadena, CA 91125, USA}
\author{Andrei Faraon}
 	\email[]{faraon@caltech.edu}
	\affiliation{Kavli Nanoscience Institute and Thomas J. Watson, Sr., Laboratory of Applied Physics, California Institute of Technology, Pasadena, California 91125, USA}
	\affiliation{Institute for Quantum Information and Matter, California Institute of Technology, Pasadena, California 91125, USA}

\date{\today}

\maketitle

{\bf Quantum emitters coupled to optical resonators are quintessential systems for exploring fundamental phenomena in cavity quantum electrodynamics (cQED) \cite{Haroche1989} and are commonly used in quantum devices acting as qubits, memories and transducers \cite{Awschalom2018}. Many previous experimental cQED studies have focused on regimes where a small number of identical emitters interact with a weak external drive \cite{Walther2006,Thompson1992,Englund2007,Lukin2022}, such that the system can be described with simple effective models. However, the dynamics of a disordered, many-body quantum system subject to a strong drive have not been fully explored, despite its significance and potential in quantum applications \cite{Kurucz2011,Diniz2011,Afzelius2010,Williamson2014}. Here we study how a large inhomogeneously broadened ensemble of solid-state emitters coupled with high cooperativity to a nanophotonic resonator behaves under strong excitation. We discover a sharp, collectively induced transparency (CIT) in the cavity reflection spectrum, resulting from quantum interference and collective response induced by the interplay between driven inhomogeneous emitters and cavity photons. Furthermore, coherent excitation within the CIT window leads to highly nonlinear optical emission, spanning from fast superradiance to slow subradiance \cite{Dicke1954}. These phenomena in the many-body cQED regime enable new mechanisms for achieving slow light \cite{Novikova2021} and frequency referencing, pave a way towards solid-state superradiant lasers \cite{Bohnet2012}, and inform the development of ensemble-based quantum interconnects \cite{Williamson2014,Afzelius2010}.}

Cavity quantum electrodynamics (cQED) offers the ability to investigate and understand the interactions between light and matter at the most fundamental level \cite{Haroche1989}. The field has enjoyed great experimental advancements in the past decades, as the rapid development of microscopic and nanoscopic devices and laser trapping techniques have revealed a diverse and rich set of phenomena \cite{Walther2006,Blais2021}. Such progress has also led to cQED’s use in quantum technology applications, including quantum information processing \cite{Duan2004,Dordevic2021}, light field manipulation \cite{Mucke2010,Englund2007}, single photon generation \cite{Keller2004}, and quantum communication \cite{kimble2008,Afzelius2010,Reiserer2015}, as the ability to change the emitters' properties with light (and vice versa) has proven to be an indispensable tool for highly controlled quantum operations.

While many works in cQED have focused on one or a few cavity-coupled emitters \cite{Thompson1992,Yoshie2004,Dordevic2021,Mucke2010,Englund2007,Keller2004,Reiserer2015,Mlynek2014,Mirhosseini2019,Evans2018}, there has been growing interest in the study of cQED with a macroscopic ensemble of emitters \cite{Norcia2018,Angerer2018,periwal2021,Blaha2020}, as the increased complexity offers deeper fundamental insights as well as expanded technological capabilities. Cavity-coupled ensembles of rare-earth ions doped in solids are an ideal platform for such a study \cite{Temnov2005,Greiner2000}, as they offer highly stable transitions in both the optical and microwave domain at cryogenic temperatures \cite{Thiel2011} and can be readily integrated into nanoscale devices \cite{Zhong2016}. In contrast to atomic gas systems \cite{periwal2021,Norcia2018}, the solid-state implementation offers the added benefit of on-chip integration for quantum applications such as high bandwidth quantum memories and transducers \cite{Businger2022,Lauk2020}. Here, the high bandwidth is necessary for frequency multiplexing in memories and high speed conversion in transducers, and is achieved as a result of the natural spectral inhomogeneity of the solid-state emitters. In order for such devices to operate efficiently, one must engineer a system with high cooperativity, which is a dimensionless figure of merit that describes the ratio between the collective coupling strength of the cavity-emitter system to dissipation, decoherence, and disorder. As improvements to material and device parameters are made towards increasing this cooperativity, it becomes critical to fully understand any associated cQED phenomena that may emerge.

\begin{figure}
    \centering
    \includegraphics[width=\linewidth]{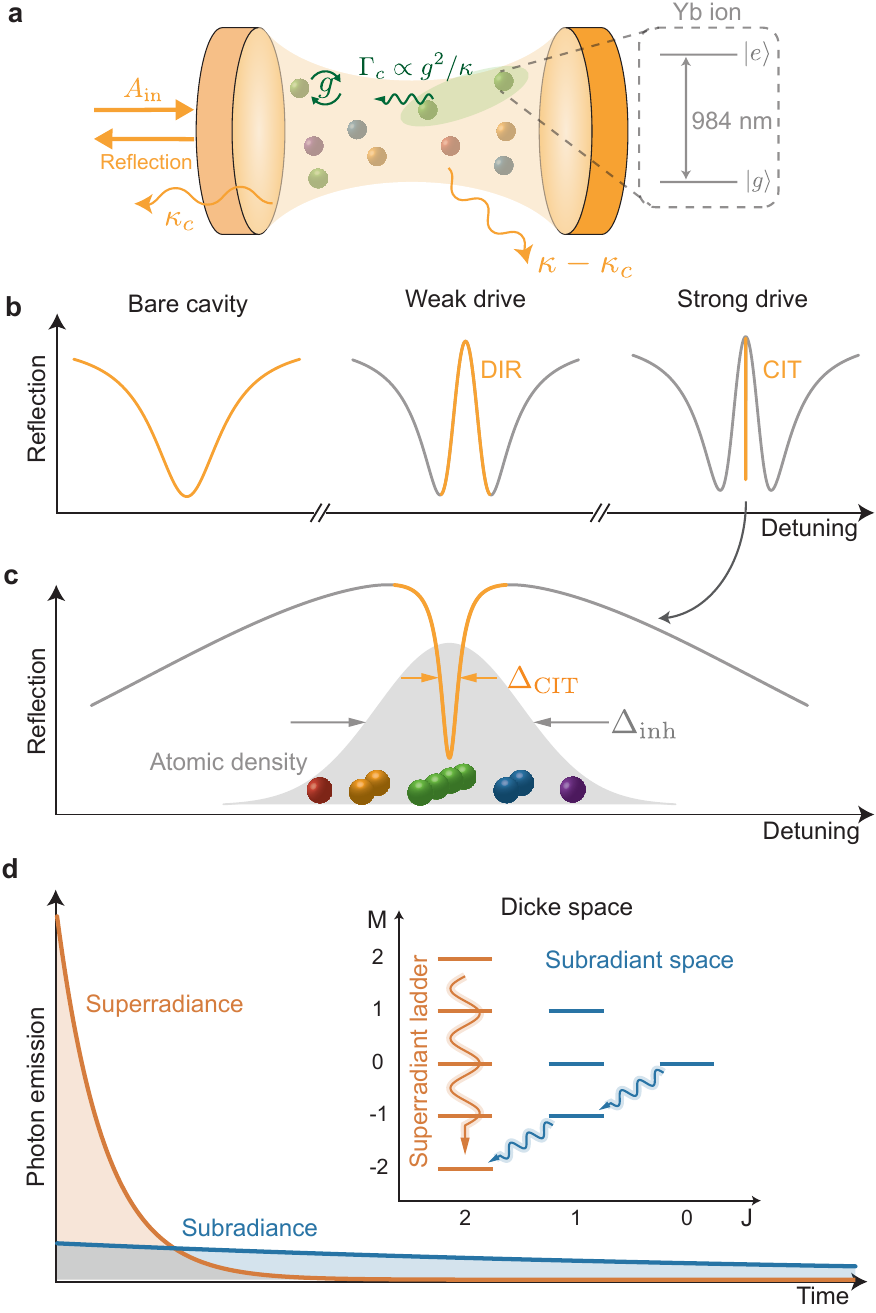}
    \makeatletter\long\def\@ifdim#1#2#3{#2}\makeatother
    \caption {{\bf Cavity QED with driven inhomogeneous emitters.} {\bf a,} Schematic description of the cavity-ion interaction. An inhomogeneous ensemble of ions is coupled to a one-sided cavity with total decay rate $\kappa$. The input field $A_\text{in}$ is coupled into the cavity with rate $\kappa_c$. The interaction strength between the cavity field and a single ion is $g$. Two spectrally indistinguishable ions have an effective cavity-mediated dissipation rate $\Gamma_c= 4g^2/\kappa$. Each Yb ion can be regarded as an effective two-level system consisting of a ground ($\ket{g}$) and excited ($\ket{e}$) state, whose transition wavelength is around 984 nm (grey dotted box). {\bf b,} Reflection spectrum of a cavity without ions showing the bare cavity resonance (left), with ions under weak drive showing DIR (middle), and with ions under strong drive showing DIR and CIT (right). {\bf c,} Zoom in of the right panel of b, showing CIT. CIT opens a transmissive window narrower than the inhomogeneous linewidth, and allows for the excitation of ions in the center of the ion distribution. {\bf d,} Schematic of superradiance (orange) and subradiance (blue) which are enhanced (suppressed) decays where emissions from ions constructively (destructively) interfere. Inset shows the Dicke space for $4$ two-level systems in the $\ket{J,M}$ basis \cite{gross1982}, where the $J=2$ manifold forms the superradiant ladder and the rest form the subradiant subspace (Methods).}
    \label{fig:fig1}
\end{figure}

In this work, we study an ensemble of approximately $10^6$ $^{171}$Yb$^{3+}$ ions embedded in YVO$_4$ coupled to a nanophotonic cavity (Fig.~1a, Extended Data Fig.~\ref{setup}), subjected to a strong driving field such that the resonant ions are excited. The relatively low spectral inhomogeneity, the strong transition dipole moment \cite{Kindem2018}, and the cavity coupling lead to a high optical cooperativity of up to 24 (Supplementary Information). This allows for strongly enhanced light-matter interactions, enabling the probing of complex collective and many-body phenomena \cite{Reitz2022}. In particular, we discover a sharp transparency window in the cavity reflection spectrum, which we call collectively induced transparency (CIT, Fig.~1b,~c). We find that the quantum interference of many inhomogeneously broadened emitters plays a critical role in producing the CIT window, mechanistically distinguishing itself from other types of transparencies \cite{Qin2020}. Taking advantage of the CIT effect, we further control the population distribution within the Dicke space \cite{Dicke1954}, which allows the observation of dissipative many-body dynamics in the form of superradiance and subradiance (Fig.~1d). The features of the observed dynamics are well explained by numerical simulations based on a many-body master equation.

\section*{Collectively Induced Transparency}
To explore cQED phenomena for a driven, inhomogenous many-body system, we first characterize the cavity-ion coupling by measuring the cavity reflection spectrum. Scanning with low laser power, the spectrum reveals broad peaks centered around the atomic resonances reaching unit reflection with about $3$ GHz width, larger than the ensemble inhomogeneous linewidth of $150$ MHz (Extended Data Fig.~\ref{ioncavitycoupling}). These peaks are known as dipole induced reflectivity (DIR), resulting from strong ion-cavity coupling (Fig.~1b, middle) \cite{Waks2006}. Specifically, in steady state under continuous driving, the cavity field $\langle a \rangle$ depends on the sum of the atomic coherences $\langle \sigma_j^-\rangle$ of individual emitters as $\langle a \rangle \propto \sum_{j=1}^{N} \langle \sigma_j^-\rangle$, where $\sigma_j^-=\ket{g}_j\bra{e}$ for the $j^{\text{th}}$ ion. As such, even if most ions remain in the ground state at weak excitation, the Yb ions still modify the internal cavity field due to the nonzero atomic coherence. This in turn influences the cavity reflectivity, leading to DIR. However, when the laser power is increased, we observe the formation of a sharp dip around the center of the DIR, which both deepens and narrows with increasing power (Fig.~2b). A Lorentzian fit to the dip gives a minimum width of $50$ MHz, and a maximum normalized depth approaching $1$ (Fig.~2d, Methods).

\begin{figure}[tp]
    \centering
    \includegraphics[width=\linewidth]{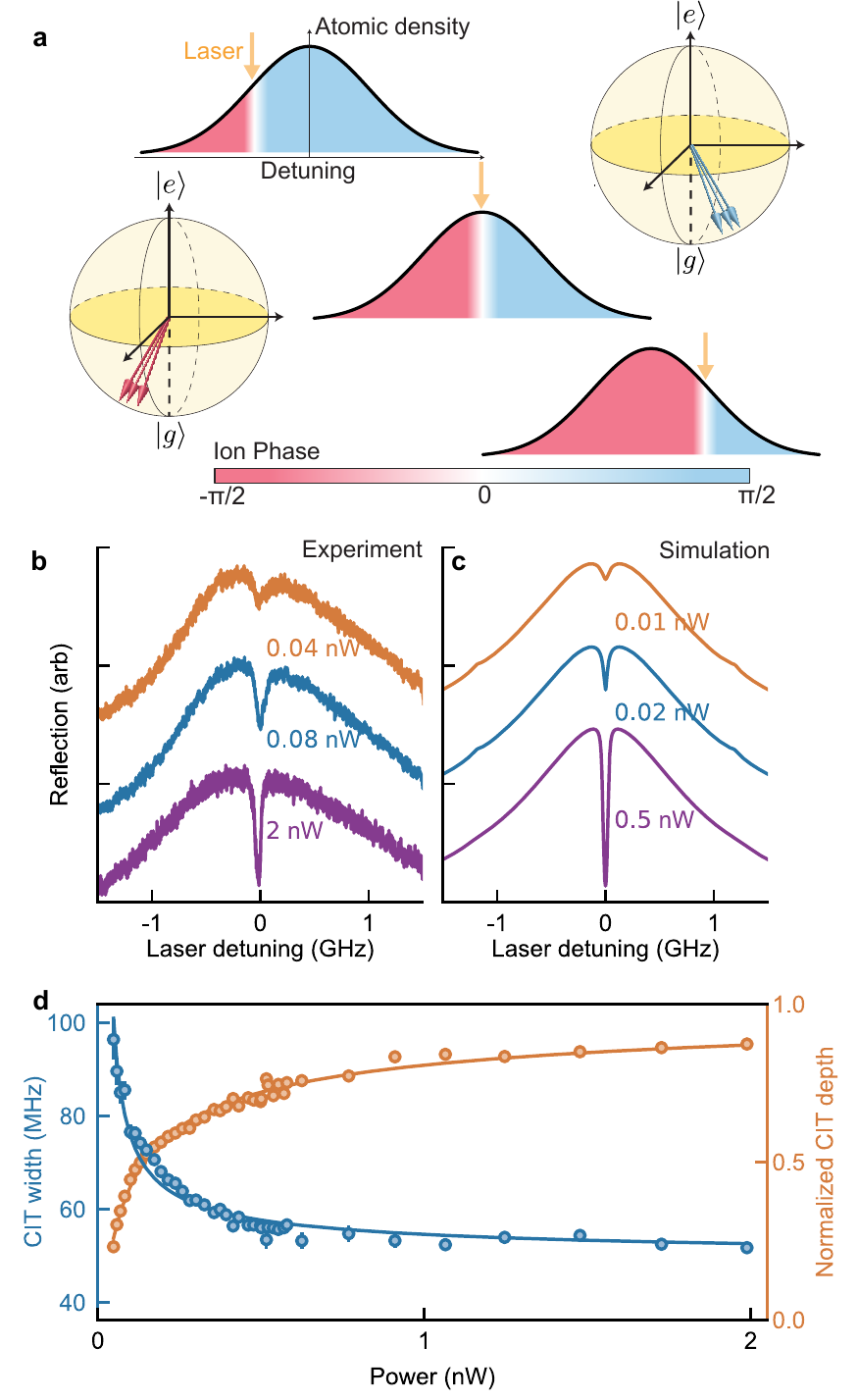}
    \makeatletter\long\def\@ifdim#1#2#3{#2}\makeatother
    \caption{{\bf Collectively induced transparency.} 
    {\bf a,} Physical origin of CIT, showing the phase distribution of atomic coherence of the inhomogeneous ensemble. The ions far-detuned from the laser frequency are out of phase on the two sides, denoted in red and blue. The resonant ions are denoted in white. The left and right Bloch spheres contain the Bloch vectors of the red- and blue-detuned ions, respectively. Three scenarios with different laser frequencies are shown. If the laser is detuned from the ensemble center (top, bottom), there is imbalance in the ion phases, while if the laser is centered (middle), the ion phases cancel and lead to CIT.
    {\bf b,} Measured cavity reflection spectra at three different powers, vertically shifted for clarity. Sharp dips due to CIT appear at the center of the DIR peaks, which get deeper and narrower with increased power. {\bf c,} Corresponding simulated cavity reflection spectra (Methods). {\bf d,} Extracted Lorentzian fit width (blue) and depth (orange) with power, error bars represent the standard errors of the fit. The CIT depth is normalized with respect to the cavity depth. Solid lines are the fits from our theoretical prediction (Methods).}
    \label{fig:fig2}
\end{figure}

We find that the origin of such a transparency window can be understood as the collective contribution of the inhomogeneous ensemble to the cavity field (Fig.~2a). For clarity the individual contributions of on- and off-resonance ions (with respect to the laser) should be considered separately. For resonant ions, strong driving saturates their steady-state populations to the completely mixed state, where both atomic inversion and coherence vanish, thus having no influence on the cavity field. In contrast, the off-resonant ions are only weakly excited, such that their atomic coherence is inversely proportional to the ion detuning $\Delta_j$, that is, $\langle \sigma_j^- \rangle \propto \Delta_j^{-1}$ (Supplementary Information). This means that ions at equal and opposite detunings are out of phase with equal amplitude, such that their pairwise contributions to the cavity field will destructively interfere. In particular, when the laser frequency is in the center of the inhomogeneous line, all of the contributions from the detuned ions cancel with each other (Fig.~2a, center). Thus, the combination of these two effects, (1) the saturation of the on-resonance ions and (2) the pairwise destructive interference of the off-resonant ions, leads to a transparency (the CIT) that emerges at the center of the inhomogeneous line (Methods). It is worth noting that CIT is unique to systems consisting of a large ensemble of emitters with an appreciable inhomogeneous broadening \cite{Jevon2021}, and does not occur for just a few emitters (Supplementary Information).

Going beyond the qualitative description, we derive an analytical expression for the width of CIT ($\Delta_\text{CIT}$) using the $N$-atom Tavis-Cummings Hamiltonian \cite{Tavis1968} under appropriate approximations (Methods):
\begin{equation}  \label{dip}
   \Delta_{\text{CIT}}\approx\frac{\Delta_{\text{inh}}}{C}\frac{1}{\left(1-C\sqrt{\frac{\gamma_s\gamma}{4g^2\mu}}\right)}
\end{equation}
where $\gamma=\frac{\gamma_s}{2}+\gamma_d$ is the total decoherence rate, comprised of the spontaneous decay rate $\gamma_s$ and the excess dephasing rate $\gamma_d$, $g$ is the single ion-cavity coupling rate, $C=~4Ng^2/(\kappa\Delta_\text{inh})$ is the ensemble cooperativity, and $\mu$ is the cavity mean photon number in the absence of ions, representing the rescaled driving laser power (Methods). The measured CIT widths and depths show excellent agreement with the predicted power dependence (Fig.~2d, Methods). Crucially, at high powers we expect the dip width to be narrowed by the ensemble cooperativity, reaching $\Delta_\text{CIT} \approx \Delta_\text{inh}/C$. Intuitively, this is because higher cooperativity leads to a larger contribution towards DIR for even a small number of imbalanced ions, effectively increasing the sensitivity to the imbalance near the ensemble center, which narrows the CIT. This indicates that if $C \gg 1$, the CIT width can be significantly narrower than the inhomogeneous broadening of an ensemble, ultimately limited by the homogeneous linewidth (Extended Data Fig.~\ref{gamma}). Given our $C$ and $\Delta_\text{inh}$, the expected minimum linewidth is $13$ MHz, narrower than the measured value of $50$ MHz. This discrepancy can be partially attributed to spectral diffusion, which effectively increases $\gamma$ and causes a breakdown of some of the assumptions made in order to derive the approximate analytical expression Eq.~\ref{dip} (Methods). To account for this, numerical simulations of the cavity reflection (without the above approximations) provides a better match to the experimental width (Fig.~2c, Methods). 

\begin{figure*}[t]
    \includegraphics[width=0.95\textwidth]{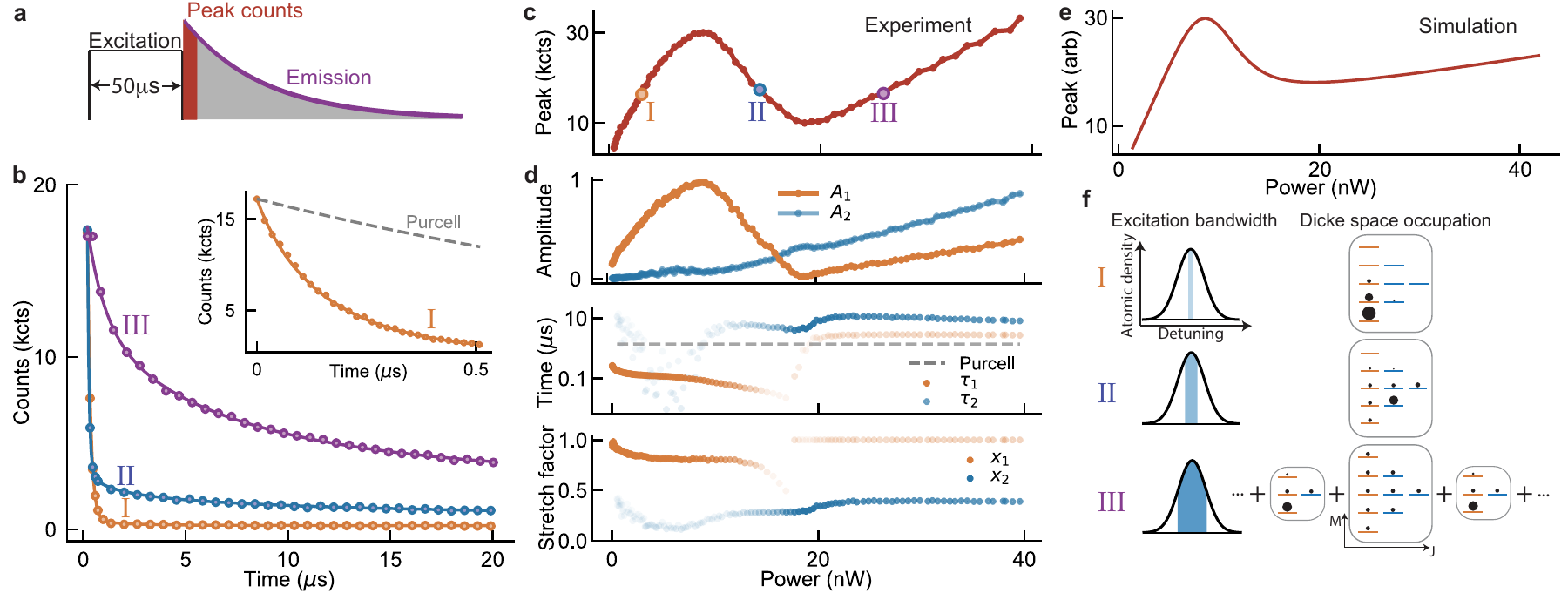}
    \makeatletter\long\def\@ifdim#1#2#3{#2}\makeatother
    \caption{{\bf Observation and analysis of dissipative many-body cavity emission.} {\bf a,} Measurement schematic. After a $50$ $\mu$s long excitation pulse, peak counts are obtained by integrating the counts within the first $128$ ns. {\bf b,} Three representative cavity emission time traces from each of the three power regimes with equal peak counts (3, 14, 26 nW). All traces are fit to a phenomenological stretched bi-exponential fit, $A_1\exp[-(t/\tau_1)^{x_1}]+A_2\exp[-(t/\tau_2)^{x_2}]+b$ (Methods). Inset: single exponential fit for the first $500$ ns of the emission excited with $3$ nW, with higher time resolution. The fitted decay lifetime of $\approx150$ ns (solid line) shows an enhancement beyond the fastest expected Purcell decay of $\approx1.4$ $\mu$s (gray dashed). {\bf c,} Peak counts with varying excitation power. The three labelled points correspond to the data shown in b. {\bf d,} Fit parameters, $A_{1,2}$, $\tau_{1,2}$, and $x_{1,2}$, of the time traces. The transparency of the data points in the decay times $\tau_{1,2}$ and stretch factors $x_{1,2}$ are weighted by their relative amplitudes $\frac{A_{1 ,2}}{A_1+A_2}$ to emphasize the significance of the parameters. In regime III (for powers greater than $20$ nW), $x_1$ is set to $1$, as there is no longer a distinct fast decay (Methods). {\bf e,} Master equation simulation of $91$ subensembles of identical ions, for a total of $N=569$ ions. Incoherently adding the peak emission of detuned subensembles qualitatively reproduces the S-curve observed in experiment (Methods). {\bf f,} Schematics of excitation bandwidth and Dicke space occupation in the three power regimes. In regime I, only a narrow bandwidth is excited, and primarily the low-excitation superradiant states are occupied. In regime II, the bandwidth increases, and the subradiant subspace becomes populated. In regime III, the bandwidth further increases, and the off-resonant subensembles get excited while leaving the on-resonance subensemble in a completely mixed state. 
    }
    \label{fig:fig3}
\end{figure*}

\section*{Dissipative many-body dynamics}

CIT enables the investigation of the rich dynamics of a driven subensemble near the transparency window, as the effect of the off-resonant ions on the cavity field is cancelled and more light is allowed to enter into the cavity. To probe the dynamics, we tune the laser to the center of the CIT, and detect the cavity emission after pulsed excitation (Fig.~3a). For state initialization, the system is driven to a non-equilibrium steady state using a long pulse (Supplementary Information). Varying the excitation power prepares the system into different initial states, followed by distinct emission dynamics (Fig.~3b). Analyzing the peak counts of the emission, we find that the trend of peak counts with power is strongly non-monotonic, forming an S-shaped curve (Fig.~3c).

To systematically characterize the observed nonlinear dynamics, we classify three power regimes (I, II, III) based on the slope of the S-curve. In regime I with low powers, the decay is predominantly fast. A characteristic $1/e$ decay time is measured to be $\approx 150$ ns, faster than the fastest expected Purcell decay of a single ion coupled to the cavity ($\approx1.4$ $\mu$s, Fig.~3b inset). In regime II, with intermediate powers, both a fast and a slow decay compared to the Purcell-enhanced rate are observed. In regime III, with higher power, a continuum of different decay lifetimes are observed, leading to a stretched exponential decay. 

To gain a microscopic understanding of this nonlinear power dependence, we use a master equation to describe driven dynamics in the presence of decoherence and dissipation. The numerical simulation of the entire inhomogeneous ensemble of $N \sim 10^6$ ions is not tractable. However, the phase cancellation in CIT negates the influence of the off-resonance ions on cavity field, which allows us to initially only consider the dynamics of the resonant ions. Additionally, we note that the cavity mediates photon exchange between ions, which triggers collective dissipation with rate proportional to $\Gamma_c=4g^2/\kappa$ (Methods). As the system dissipation is dominated by $\Gamma_c$, a smaller number of ions that sit within a spectral window whose width is about $\Gamma_c$ can be treated as indistinguishable ions. To this end, we first simulate a small-scale homogeneous ensemble. Specifically, we study a toy model of 6 identical ions whose dynamics can be described using the Dicke states, the coupled basis defined for indistinguishable two-level systems \cite{Dicke1954}. As shown in Fig.~1d and Extended Data Fig.~\ref{decayrates}, vertical decays between the Dicke states are enhanced and superradiant, and diagonal decays are suppressed and can only decay through individual dissipation channels, which we call subradiance (see Methods for details including semantics). To effectively capture the existence of multiple decay rates among the various Dicke states, as well as a clear separation between fast (superradiant) and slow (subradiant) decays, we employ a phenomenological stretched bi-exponential fit and extract the relevant fit parameters, which also clearly reveals the presence of the distinct three regimes discussed earlier (Fig.~3d, Methods).

By simulating this system's dynamics using the master equation, we find that the peak emission is a good indicator for the population distribution of the Dicke states prepared by the drive (Methods). The simulated peak emission matches the trend measured in regimes I and II, where distinct temporal dynamics are attributed to decays from different parts of the Dicke space (Extended Data Fig.~\ref{sixmaster}). In regime I we attribute the fast decay to superradiance, dominantly from the collective dissipation within the superradiant ladder, as we expect to have populated only the low-excitation superradiant states within a narrow bandwidth of the ensemble (Fig.~3f, top). From the measured fast decay rate, we estimate the number of ions participating in superradiance to be on the order of $\sim50$ (Supplementary Information). With increased power, we expect that the system climbs up the superradiant ladder and reaches Dicke states with larger decay rates, leading to even faster emission. This is consistent with the observed trend of decreasing $\tau_1$ in regime I as shown in Fig.~3d. At even higher powers, strong driving of the superradiant ladder allows for significant population to diffuse into the subradiant space via decoherence processes, resulting in the slow decay observed in regime II (Fig.~3f, middle) \cite{Cipris2021,Glicenstein2022}. Populating multiple dark subradiant states exhibiting different decay rates manifests as a stretched exponential decay in the emission dynamics.

In regime III a completely mixed state of equal population in each Dicke state can be reached (Fig.~3f, bottom). Further increasing the power excites more of the off-resonance ions, while the subensemble of the on-resonance ions addressed in regimes I and II remains in the completely mixed state. This leads to the emergence of intermediate decays, departing from the homogeneous Dicke subensemble picture, which suggests that a wider excitation bandwidth at high powers should be considered in numerical simulations. To this end, we simulate a larger number of emitters by including a Lorentzian distributed ensemble of ions with experimental inhomogeneous linewidth (Methods). Specifically, the dynamics of each subensemble is computed separately and incoherently added, by assuming that the subensembles are effectively non-interacting (Extended Data Fig.~\ref{incadd}). The emergence of the upturn of the peaks counts at high powers is reproduced by the simulation (Fig.~3e), consistent with the experimental observation in regime III.

\section*{Control over coherent emission}
\begin{figure}
    \centering
    \includegraphics[width=\linewidth]{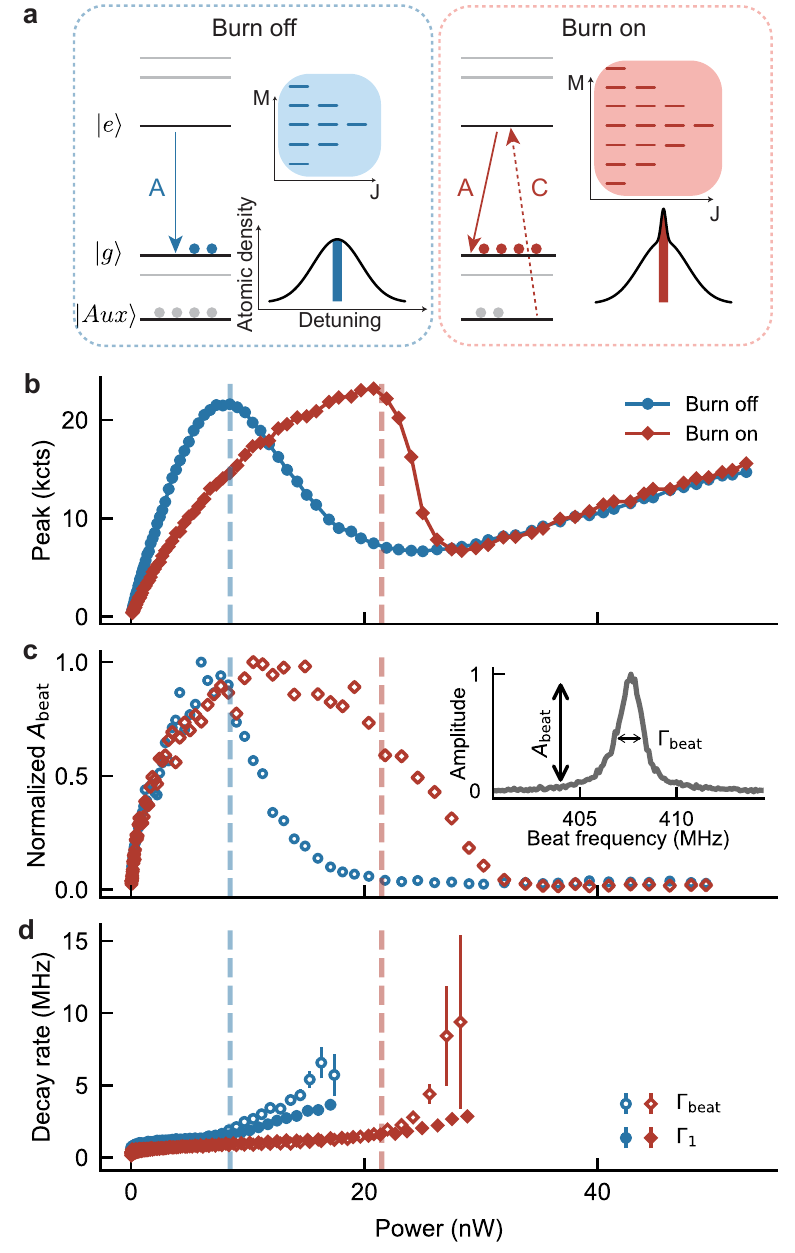}
    \caption{{\bf Control and characterization of dissipative many-body dynamics via hole-burning} {\bf a,} Schematic representations of hole-burning (left: no hole burning, right: with hole burning). The C transition connects another state $\ket{Aux}$ in the ground-state manifold to the excited state $\ket{e}$. Optical pumping on the C transition transfers population from $\ket{Aux}$ to $\ket{g}$, introducing an anti-hole in the original distribution of the A transition. This increases the local ion density of the A transition at the probe frequency (colored thin bars), and expands the Dicke space (shaded areas). See Methods for the detailed description of the A, C transitions and the $\ket{Aux}$ state.
    {\bf b,} Peak emission counts measured as a function of excitation laser power with (red) and without (blue) hole burning. S-curve shifts towards higher power when the burning is on. {\bf c,} Normalized beat note amplitude $A_\text{beat}$ for burning off (blue) and on (red). The two data sets are separately normalized, as the amplitude strongly depends on the local oscillator polarization, which can vary between experiments. Inset shows the experimental characterization of width $\Gamma_\text{beat}$ and amplitude $A_\text{beat}$ of a beat signal from a heterodyne measurement (Methods). {\bf d,} Single exponential fit $\Gamma_{\text{1}}$ (filled markers) from the time-dynamics measurement and extracted width $\Gamma_{\text{beat}}$ from a Lorentzian fit (open markers) from coherence measurements for burning off (blue) and on (red). Error bars represent standard errors of the fits. In b-d, vertical dotted lines indicate the turning points between regimes I and II.}
    \label{fig:fig4}
\end{figure}
To demonstrate control over the dissipative many-body dynamics, and to show further evidence of the beyond-single atom nature of the cavity emission, we first modify the decay dynamics by changing the number of ions. Specifically, we tune the number of ions resonant with our excitation laser $N_\text{res}$ via optical hole burning, and then observe the changes in the S-curve (Fig.~4a). Upon increasing $N_\text{res}$, the S-curve shifts towards higher power in regimes I and II, along with an increase in the maximum of peak counts (Fig.~4b). The shift implies the formation of a larger superradiant ladder when the number of local homogeneous ions increases, which requires more excitation power to optically pump into the subradiant subspace (Supplementary Information). However, regime III is observed to be largely insensitive to a change in $N_\text{res}$, as indicated by the overlap of the S-curves, since spectral hole-burning only changes the population locally in frequency without affecting the number of detuned ions as illustrated in Fig.~4a.

We provide additional evidence of the traversal of the Dicke space by measuring the coherence of the emission via heterodyne detection (Methods). The beat-note between the emission and the excitation laser provides the rate and amount of coherent decay through its width $\Gamma_{\text{beat}}$ and amplitude $A_{\text{beat}}$, respectively (Fig.~4c, inset). We first note the rise of $A_\text{beat}$ in regime I, indicating the increase in population of the superradiant ladder. Later, $A_\text{beat}$ decreases in regime II, corresponding to the incoherent coupling to the subradiant subspace. Finally in regime III, $A_\text{beat}$ vanishes because of the absence of coherent decay, as the population has undergone incoherent processes to reach the completely mixed state.

Since only decays within the superradiant ladder are coherent with respect to the excitation, $\Gamma_\text{beat}$ represents the rate of superradiance within the superradiant ladder. For comparison, we extract the fast decay part of the time dynamics of photon emission as a single exponential with the rate $\Gamma_{1}$, which captures all of the enhanced decays from both superradiant and subradiant subspaces (Fig.~4d, filled markers). The comparison between $\Gamma_\text{beat}$ and $\Gamma_1$ can then be used to evaluate the relative decay contributions from the two subspaces. We find that $\Gamma_{1}$ overlaps with $\Gamma_{\text{beat}}$ for low powers, confirming that all of the fast decays in regime I are within the superradiant ladder. However, entering regime II, $\Gamma_{1}$ deviates from $\Gamma_{\text{beat}}$ (for powers beyond dashed lines in Fig.~4d). This observation of $\Gamma_1 < \Gamma_\text{beat}$ in regime II indicates that $\Gamma_1$ also includes some incoherent decays slower than $\Gamma_\text{beat}$, which point to the enhanced decays within the subradiant subspace (Extended Data Fig.~\ref{decayrates}a).

Lastly, we have also performed another measurement by varying the frequency of the probe laser to control the number of driven ions and observed that the nonlinear S-curve shifts along the expected direction (Extended Data Fig.~\ref{vshape}). We have also confirmed that all of the experimental findings for CIT and sub- and superradiance are reproducible and consistent with our theoretical predictions, independent of the choice of the optical emission lines between the A, E, and I transitions (Extended Data Fig.~\ref{AIcomp}). All of these complementary experiments lend strong support to our microscopic understanding and control of an inhomogeneous ensemble.

\section*{Discussion and outlook}
In this work we have investigated the spectral response and open quantum dynamics of a large cQED system, revealing a sharp CIT and highly nonlinear, dissipative many-body dynamics. Notably, the CIT width is found to be narrowed by cooperativity, indicating that improvements in fabrication and material properties towards increasing cooperativity can lead to much narrower transparencies, potentially useful as frequency references. The sudden cavity phase shift across the CIT (Extended Data Fig.~\ref{phaseshift}) can provide a novel mechanism to achieve optical nonlinearities and the storage of light \cite{Novikova2021}. In particular, we demonstrate a proof-of-principle optical switch using CIT, and posit that with further optimization a fast, high contrast two-port optical switch can be realized (Extended Data Fig.~\ref{citresponse}).

Further, the observed optical superradiance and subradiance represent a key step towards enabling narrow linewidth superradiant lasers \cite{Bohnet2012} and long-lived subradiant memories \cite{Shen2022,Ferioli2021} in solid-state, while the control over the population of the Dicke space opens the door for dissipation-based engineering of state preparation \cite{Cirac2009,Kastoryano2011}. In addition, operating with a detuned cavity may allow the probing of coherent photon-mediated interaction between the ions (Supplementary Information), opening new possibilities for studying coherent spin exchange effects and quantum simulations \cite{Norcia2018,Swan2021} in a solid-state platform. Finally, the improved understanding in this regime of collective and many-body cQED phenomena informs the development of high-cooperativity solid-state quantum memories and transducers \cite{Williamson2014,Afzelius2010}.

\bibliographystyle{naturemag}
\bibliography{apssamp}

\clearpage

\section*{Methods}
\subsection*{Device}
The substrate is a 3$\times$3$\times$0.5 mm piece of $^{171}$Yb$^{3+}$:YVO$_4$ (a $\times$ a $\times$ c), measured to have a Yb doping concentration of 86 parts-per-million using glow discharge mass spectrometry \citemethods{Bartholomew2020}. The device is fabricated directly in the substrate using focused ion-beam milling. The optical cavity is formed by periodic grooves milled into a triangular nanobeam waveguide, with a slight aperiodicity in the center which forms a defect creating the cavity mode. A 45-degree angled coupler couples the light from free-space to the waveguide with an efficiency of $\approx$ 25\%. Further details on the device fabrication can be found in \cite{Zhong2016}.

Based on the concentration of Yb ions and the cavity volume, we estimate that about $N\approx7\times 10^5$ ions are coupled to the cavity with varying coupling strengths (Supplementary Information). The cavity is tuned into resonance with the $^2$F$_{7/2}$ to $^2$F$_{5/2}$ transition of Yb around $\lambda=984$ nm using nitrogen gas condensation. The large cavity linewidth ($\kappa=2\pi\times44$ GHz) covers all three transitions aligned along the cavity polarization (labelled as A, E, I, Extended Data Fig.~\ref{ioncavitycoupling}a). The narrow optical inhomogeneous linewidths ($\Delta_{\text{inh}}=2\pi\times150$ MHz) compared to the separation between those transitions (a few GHz) enables each transition to be addressed as independent two-level systems. Because of this, in the main text we have focused primarily on the A transition for simplicity. The nanoscale cavity allows for tight confinement of the electromagnetic field, resulting in a small mode volume of about a cubic wavelength ($\approx 1(\lambda/n)^3$, where $n$ is the refractive index). In conjunction with the relatively strong dipole moment of Yb in YVO$_4$ \cite{Kindem2018,Thiel2011}, these factors enable high vacuum Yb ion-cavity coupling $g$ up to $\approx2\pi\times35$ MHz, leading to a large collective ensemble cooperativity $C = 4Ng^2/(\kappa \Delta_\text{inh}) > 1$ in the optical regime. Considering the distribution of $g$, we obtain the root mean square of $g$ as $\sqrt{\langle g^2 \rangle}\approx2\pi\times10.6$ MHz (Supplementary Information). Using this, we extract $C\approx 12$ for the A, E transitions and $C \approx 24$ for the I transition which has twice the population as it connects degenerate doublets, in good agreement with expectation from system parameters (Supplementary Information).

\subsection*{Experimental setup}
The optical setup for the experiments is shown in Extended Data Fig.~\ref{setup}. Not all parts of the setup are used in all of the measurements. The lasers addressing transitions A and C are both Toptica DL Pro, tunable around $980$ nm. Both lasers can be frequency locked (not shown in Extended Data Fig.~\ref{setup}) to a stable reference optical cavity using the Pound-Drever-Hall method, and we measure a laser linewidth of approximately $600$ Hz over $10$ $\mu$s using the delayed homodyne method. The lasers can also be frequency-swept by modulating the internal piezo-electric actuator. In this mode the laser is free-running, where the linewidth is measured to be $50$ kHz, with a slower drift in the center frequency on the order of a few MHz over tens of seconds. A Thorlabs S130 photodiode power sensor is used to measure excitation powers. The actual powers that reach the cavity are calibrated by measuring round-trip losses from the cavity. We measure approximately $10$\% of the light reaches the device, including the angled coupler efficiency of $\approx25$\%. However, we note that due to slight differences between measurements of the laser polarization and device coupling, there is likely slight discrepancies in all calibrated powers.

Acousto-optic modulators (AOMs) are used to gate optical pulses for pulsed measurements. Two AOMs are used in series for the probe laser giving an extinction of $\approx$ $100$ dB. The light is sent to the device via an optical circulator (Precision Micro-Optics), and focused onto the angled coupler with an aspheric lens doublet, which is mounted on a 3-axis piezo nano-positioner stack (Attocube) for fine alignment. The device itself is mounted on the mixing chamber plate of a Bluefors dilution refrigerator with a base temperature of around 40 mK with no external magnetic field applied. The reflected signal from the circulator is sent to a superconducting nanowire single photon detector (SNSPD) held at $900$ mK, and photon counts are time-tagged with a Swabian Time Tagger 20. A gating AOM is used before the SNSPD to selectively attenuate the intense reflected input pulses.

The coherence measurements of the cavity emission were taken by splitting off part of the input laser as a local oscillator to beat with the emission \citemethods{Shcherbatenko16}. The beat signal was subsequently detected by the SNSPD and Fourier transformed to obtain the power spectra. All of the RF drives used to drive the AOMs were phase synchronized. To maximize the signal-to-noise ratio, it is desirable to integrate for a long time. However, long integration time requires phase stability of all of the parts of the experiment, particularly the fibers. Due to this, we found that an integration time of $1$ second provides sufficient signal-to-noise ratio while maintaining the phase stability. To further improve the signal-to-noise ratio, we repeatedly integrated the signal for $1$ second and averaged the power spectra. An example of the power spectra is shown in Fig.~4c inset, where the beating frequency is around $408$ MHz, given our chosen frequency difference between our probe and local oscillator.

\subsection*{Theoretical model}
To model our system, we first regard each ion as a two-level system ($\ket{e}$ and $\ket{g}$ as labelled in Extended Data Fig.~\ref{ioncavitycoupling}a) and consider the Tavis-Cummings Hamiltonian in the laser frame \cite{Tavis1968}:
\begin{align} \label{h}
    H = \Delta_c a^{\dagger}a + \frac{1}{2}\sum_{j=1}^{N} \Delta_j\sigma_j^z +& g\sum_{j=1}^{N} (a^{\dagger}\sigma_j^- + \sigma_j^+a) \nonumber \\
    &-i\sqrt{\kappa_{c}}A_\text{in}(a^{\dagger}-a).
\end{align}
Here, $a$ is the bosonic cavity field operator, $\sigma_j^{\pm}$ and $\sigma_j^z$ are the spin ladder operators and the Pauli-Z operators describing the atomic coherence and inversion of the $j^{\text{th}}$ ion, respectively. $\Delta_c$ is the cavity-laser detuning, $\Delta_j$ is the $j^{\text{th}}$ ion-laser detuning, $g$ is the ion-cavity coupling rate, and $\sqrt{\kappa_c}A_{\text{in}}$ is the excitation field strength that enters the cavity, where $\kappa_c$ is the input coupling rate and $A_{\text{in}}=\sqrt{\frac{P_\text{in}}{\hbar \omega}}$ is related to the input laser power $P_{\text{in}}$ at frequency $\omega$. Note that here we consider homogeneous $g$ for simplicity, see Supplementary Information for discussion on inhomogeneous $g$.

To model the cavity reflection spectrum and CIT, we use the above Hamiltonian and derive the equations of motion for $a$, $\sigma_j^-$ and $\sigma_j^z$ in the Heisenberg picture:
\begin{equation} \label{a}
    \dot{a} =  - (i\Delta_c+\frac{\kappa}{2})a-ig\sum_{j=1}^{N} \sigma_j^- - \frac{\kappa}{2}\sqrt{\mu}
\end{equation}
\begin{equation} \label{sigma-}
    \dot{\sigma}_j^- = -(i\Delta_j+\gamma)\sigma_j^- +ig\sigma_j^za
\end{equation}
\begin{equation} \label{sigmaz}
    \dot{\sigma}_j^z = 2ig(a^{\dagger}\sigma_j^- - \sigma_j^+a)-\gamma_s(1+\sigma_j^z)
\end{equation}
where we have introduced the operators' corresponding dissipation terms, cavity decay rate $\kappa$, total atomic decoherence rate $\gamma$, and spontaneous emission rate $\gamma_s$. We additionally use $\mu\approx\frac{4\kappa_c}{\kappa^2}A_\text{in}^2$, which is the cavity mean photon number in the absence of ions, representing the rescaled driving laser power (Supplementary Information).

Meanwhile for modelling the dynamics, we introduce the dissipative mechanisms through the Lindblad operators:
\begin{equation}
        \mathcal{L}_{\text{cav}}=\kappa(a\rho_t a^{\dagger} - \frac{1}{2}a^{\dagger}a\rho_t - \frac{1}{2}\rho_t a^{\dagger}a)
\end{equation}
\begin{equation}
        \mathcal{L}_{\text{em}}=\gamma_s\sum_{j=1}^N (\sigma_j^- \rho_t \sigma_j^+ - \frac{1}{2}\sigma_j^+ \sigma_j^- \rho_t - \frac{1}{2}\rho_t \sigma_j^+ \sigma_j^-)
\end{equation}
\begin{equation}
        \mathcal{L}_{\text{deph}}=\gamma_d\sum_{j=1}^N (\sigma_j^z\rho_t \sigma_j^z - \rho_t)
\end{equation}
where $\mathcal{L}_{\text{cav}}$ is the cavity dissipation, $\mathcal{L}_{\text{em}}$ is the local spontaneous emission, $\mathcal{L}_{\text{deph}}$ is the local dephasing, and $\rho_t$ is the total density operator consisting of cavity field and the atoms. As we are in the bad cavity regime where $\kappa$ is much larger than all other system rates, the cavity mode is adiabatically eliminated, which changes $\mathcal{L}_\text{cav}$ to:
\begin{equation} \label{Lcol}
 \mathcal{L}_{\text{col}}=\Gamma_c(J^-\rho J^+- \frac{1}{2}J^+J^-\rho -\frac{1}{2} \rho J^+J^-)
\end{equation}
where $J^\pm=\sum_{j=1}^N\sigma_j^\pm$ is the collective atomic coherence and $\Gamma_c = 4g^2/\kappa$ is the Purcell-enhanced decay rate of a single ion. Similarly, the cavity mode is eliminated from the Hamiltonian (Supplementary Information) giving
\begin{equation}
    H_{\text{at}} \approx \frac{1}{2} \sum_{j=1}^{N} \Delta_j\sigma_j^z - \sum_{j=1}^{N}g\mu(\sigma_j^+ + \sigma_j^-).
\end{equation}

$\mathcal{L}_{\text{em}}$ and $\mathcal{L}_{\text{deph}}$ can be reduced to the many-body atomic density operator $\rho$, obtained by taking the trace over the cavity field subspace. Using these we solve the following master equation:
\begin{equation} \label{mastereq}
    \dot{\rho}=-i[H_\text{at},\rho]+\mathcal{L}_{\text{col}}+\mathcal{L}_{\text{em}}+\mathcal{L}_{\text{deph}}.
\end{equation}

\subsection*{Derivation of analytical expression for CIT}
Using the input-output formalism, $A_\text{out} = \sqrt{\kappa_c}a + A_\text{in}$, we first obtain the cavity reflection
\begin{equation}
    R=\left|\frac{\langle A_\text{out}\rangle}{\langle A_\text{in}\rangle}\right|^2=\left|\frac{2\kappa_{c}}{\kappa\sqrt{\mu}}\langle a \rangle+1\right|^2.
\end{equation}

To get a neat analytical expression, we first assume a Lorentzian distribution of ions, and additionally make the following assumptions:
\begin{subequations} \label{AS}
\begin{align}
&\text{High cooperativity:} \nonumber \\ 
  &C\gg1\\ 
  &\text{Intermediate power:} \nonumber \\ 
  &  \left(\frac{\Delta_\text{inh}}{4g}\right)^2\frac{\gamma_s}{\gamma}\gg \mu \gg \frac{\gamma\gamma_s}{4g^2}  \\
  &\text{Appreciable inhomogeneity and good coherence:} \nonumber \\ 
  & \frac{\Delta_\text{inh}}{\gamma}\gg C 
\end{align}
\end{subequations}

With the above conditions, Eqs. \ref{a}-\ref{sigmaz} are solved in the steady state. We find that the cavity reflection $R$ as a function of laser frequency $\omega_L$ near the center of the ensemble $\omega_0$ has a Lorentzian profile:
   \begin{equation} \label{Rlor}
     R=1-\frac{A}{\left(\frac{\Delta_{\text{CIT}}}{2}\right)^2+(\omega_L-\omega_0)^2}
   \end{equation}
where $A=\frac{\kappa_c}{\kappa}(\frac{\Delta_{\text{inh}}}{C})^2\left(\frac{C\Delta_{\text{CIT}}}{\Delta_{\text{inh}}}-\frac{\kappa_c}{\kappa}\right)$ and $C=~\frac{4Ng^2}{\kappa\Delta_{\text{inh}}}$ (Supplementary Information) and
\begin{equation} \label{eqwidth}
     \Delta_{\text{CIT}}=\frac{\Delta_{\text{inh}}}{C}\frac{1}{\left(1-C\sqrt{\frac{\gamma_s\gamma}{4g^2\mu}}\right)}.
\end{equation}

The Lorentzian dip given by Eq.~\ref{Rlor} is the observed CIT dip, with width  $\Delta_{\text{CIT}}$. We further define the normalized depth $\eta_{\text{CIT}}$
\begin{equation} \label{eqdepth}
\begin{split}
    \eta_{\text{CIT}}&=\frac{A}{\left(\frac{\Delta_{\text{CIT}}}{2}\right)^2\eta_\text{bare}}\\
    &=\frac{1}{(1-\frac{\kappa_c}{\kappa})}\left(1-C\sqrt{\frac{\gamma_s\gamma}{4g^2\mu}}-\frac{\kappa_c}{\kappa}\left(1-C\sqrt{\frac{\gamma_s\gamma}{4g^2\mu}}\right)^2\right)
\end{split}
\end{equation}
which is the amplitude of this Lorentzian dip normalized by the bare cavity depth $\eta_\text{bare}=1-\left(1-\frac{2\kappa_c}{\kappa}\right)^2$. In the limit of high power ($\mu$), $\eta_{\text{CIT}}$ approaches $1$, where the absolute reflectivity will ultimately be limited by the bare cavity reflectivity. Hence if the cavity is critically coupled ($\frac{\kappa_c}{\kappa}=0.5$), zero reflection, or full transparency can be realized.

While the analytical expressions above give the intuition behind CIT (Supplementary Information), an arbitrary distribution can be numerically solved without making the assumptions listed in Eq.~\ref{AS}, which is how the results in Fig.~2c are obtained. This gives CIT widths closer to the experimental values. Note that the power used in simulation in Fig.~2c is four times smaller than the experiment in Fig.~2b, which is attributed to discrepancy of the realistic distribution of ions and power calibration errors in experiment. Specifically, we found that making the simulated ion distribution imperfect or asymmetric resulted in requiring more power to effectively reach the high power regime, where the CIT width reaches its minimum.

\subsection*{Master equation simulations of dynamics}
For modelling the dynamics, we solve the master equation in Eq.~\ref{mastereq} using QuTIP (Supplementary Information). However, the full master equation simulation of our large ensemble is intractable. To this end, we make use of the fact that in this bad cavity limit, the cavity dissipation turns into collective emission proportional to $\Gamma_c$ as in Eq.~\ref{Lcol}. Here, $\Gamma_c$ describes the cavity-mediated collective dissipation rate among ions, defining an effective spectral bandwidth within which the ions are considered to be indistinguishable (Supplementary Information). Hence, we simulate a mesoscopic, homogeneous ensemble to aid in the qualitative understanding of our system dynamics.

In simulating our system, we must first establish a connection between experimental measurements and simulatable quantities. We note that the peak counts reflects the cavity population at the end of the excitation pulse (Supplementary Information). In the fast-cavity regime, and in the absence of an input field, the cavity population depends on the atomic states as $\langle a^\dagger a \rangle = \Gamma_c \langle J^+ J^- \rangle$, where $\langle J^+J^- \rangle$ can be written as:

\begin{equation} \label{cor}
    \langle J^+J^- \rangle = \underbrace{\vphantom{\sum_{i\neq j}^N \langle \sigma_i^+ \sigma_j^- \rangle_\text{Correlation}}\sum_{i=1}^N \langle \sigma_i^+ \sigma_i^- \rangle}_\text{Individual} + \underbrace{\sum_{i\neq j}^N \langle \sigma_i^+ \sigma_j^- \rangle}_\text{Correlation}. 
\end{equation}

Here the first term is the sum of the emission of individual ions, and the second term represents the correlation between different ions. We simulate different parts in Eq.~\ref{cor} with a toy model of 6 identical ions with experimental coupling and dissipation rates (Extended Data Fig.~\ref{sixmaster}a). The trend of $\langle J^+J^- \rangle$ indeed qualitatively matches the experimental observations for regimes I and II. The initial increase of $\langle J^+J^- \rangle$ is due to the build-up of positive correlations, or superradiance. With higher power, the correlations decrease, due to an increase of population in the subradiant subspace. This is substantiated by the evolution of the population in the superradiant and subradiant subspaces with power (Extended Data Fig.~\ref{sixmaster}b, c). We note that an increase then decrease of emission can be associated with the saturation of the coherence, also seen with just a single emitter. However, we find that the underlying mechanism for our observation with dense inhomogeneous emitters is fundamentally different from the above phenomenon (see Supplementary Information for details).

\subsection*{Modelling regime III}
While the modelling of a small, homogeneous ensemble qualitatively captures the experimental behavior in regimes I and II (Extended Data Fig.~\ref{sixmaster}a), regime III cannot be modelled in this way. To this end, we include some frequency inhomogeneity to our model to capture the fact that as we increase power, we increase our excitation bandwidth, and thus excite more ions detuned from the laser. In order to incorporate more ions in our simulation, we first exploit the permutational symmetry of identical particles using Permutational Invariant Quantum Solver (PIQS, Supplementary Information) to decrease our computation time, allowing upwards of $30$ identical ions to be readily simulated.

Additionally to incorporate inhomogeneity, we would ideally like to approximate sufficiently detuned ions as separate ensembles whose contribution to the cavity population $\langle a^{\dagger}a\rangle$ can be incoherently summed. To this end we compare two cases with $7$ ions in Extended Data Fig.~\ref{incadd}. One case is simulating the full system of $7$ ions, with $2$ ions detuned by $5$ MHz. Another case is the incoherent addition of $5$ ions on resonance and $2$ ions detuned by $5$ MHz, where each system is solved separately and the peak emission summed after. While there is an offset in the values of the peak emission at certain powers, the qualitative behavior remains the same.

Combining the above two assumptions, we simulate an inhomogeneous ensemble of ions following a Lorentzian distribution. We indeed qualitatively reproduce the experimentally observed behavior in regime III, where the excitation of off-resonant ions leads to the increase of peak emission at high powers, giving rise to the nonlinear S-shaped profile.

\subsection*{Data fits}
\subsubsection*{CIT widths and depths}
As the distribution of ions in our experimental system is approximately Lorentzian, based on Eq.~\ref{eqwidth} and Eq.~\ref{eqdepth}, we use the following functions to fit the power dependent CIT width and the depth
\begin{equation} \label{eqwidthfiit}
     \Delta_{\text{CIT,fit}}=\frac{p_1}{1-\frac{p_2}{\sqrt{P}}}
\end{equation}

\begin{equation} \label{eqdepthsfit}
    \eta_{\text{CIT,fit}}=p_4\left(1-\frac{p_2}{\sqrt{P}}-p_3\left(1-\frac{p_2}{\sqrt{P}}\right)^2\right)
\end{equation}
where $p_{1,2,3,4}$ are free fitting parameters and $P$ is the excitation power. The fit parameters are left floating as the purpose of these fits are to validate the analytically derived power scaling, which contains some approximations that may make it inexact in certain regimes. This is already apparent in the discrepancy of the minimum CIT width, where the analytical value is a few times smaller than the experimental and numerically simulated values.

Regardless, we find that the extracted fit parameters from Fig.~2d and Extended Data Fig.~\ref{AIcomp} are physically reasonable based on our system parameters, for both the A and I transitions. First $p_1$, representing the minimum CIT width, is fit to $42(36)$ MHz for the A(I) transitions. $p_2$, the prefactor to the excitation power, is fit to 0.08(0.25), where the larger value for the I transition reflects both the larger cooperativity and dephasing. The analytical expression of $p_2$ is $C\sqrt{\frac{\hbar\omega\gamma_s\gamma\kappa^2}{g^2\kappa_c}}$, and plugging in system parameters we obtain about $0.07(1.4)$, accounting for optical losses and the factor of $4$ discrepancy found in the numerical simulations. We attribute the discrepancy of $p_2$ for the I transition to an overestimation of the dephasing rate, which we assumed to be a hundred times worse than the A transition. $p_3$ is the extracted fit for the cavity in-coupling ratio $\kappa_c/\kappa$, fit to $0.3(0.1)$, a good match to the estimated value of $\kappa_c/\kappa\sim0.2$ measured in similar devices. Correspondingly, $p_4$ is fit to $1.2 (1.1)$, consistent with its analytical expression $\frac{1}{(1-\kappa_c/\kappa)}$.

We note that the measured CIT depths in Fig.~2d and Extended Data Fig.~\ref{AIcomp} are normalized against the bare cavity depth, determined by $\kappa_c$. For the experiment data, we set the cavity resonance minimum to be $0$ (which we take to be the minimum of the edge of the DIR, since the cavity is broad), and the DIR maximum to be $1$. This is done to eliminate the background counts of reflected light that do not enter the cavity. 

\subsubsection*{Decay fits}
To characterize the power-dependent, non-single exponential decay profiles in Fig.~3, we employ the following phenomenological stretched bi-exponential fit
\begin{equation} \label{fit}
    y(t)=A_1\exp[-(t/\tau_1)^{x_1}]+A_2\exp[-(t/\tau_2)^{x_2}]+b
\end{equation}
with a fast stretched exponential decay with time constant $\tau_1$, amplitude $A_1$, and stretch factor $x_1$ and a slower stretched exponential decay with time constant $\tau_2$, amplitude $A_2$, stretch factor $x_2$, and background $b$ (Supplementary Information). The fit parameters (Fig.~3d) reflect the distinct decay behaviors in each of the three regimes consistent with the observations in Figs.~3b and 3c. In particular, we see a clear transition in the fitted decay time from superradiance to subradiance at around $20$ nW of power, as there is an emergence of slow decay ($\tau_2$) and increase of $\tau_1$. Further details on the justification of the fitting function are provided in the Supplementary Information.

To capture both the fast decay (which requires fine timing resolution at the nanosecond level) and the slow decay (which requires data out to $100$s of microseconds after the excitation), we employ two different data taking methodologies. To first capture the fast decay, we zoom into the first few microseconds of the decay with $1$ ns resolution. This allows us to fit the decay to a stretched exponential in regimes I and II. At the same time, a separate data set with a timing resolution of $128$ ns is taken such that we can probe out to longer timescales. We use this data set to fit the slow decay in blue. However in regime III, as shown in Fig.~3b, the decay is smoother without a clear distinction between fast and slow decay. Because of this we only use the $128$ ns timing resolution dataset, and force $x_1=1$ as here the fast decay simply samples the fastest decay in the smooth, multi-exponential profile.

\subsection*{Dicke states}
The Dicke states can be described in the $|J, M\rangle$ basis, with $J$ = [$N/2$, $N/2-1$, ...] ($J\geq0$) and $M$ = [$-J$, $-J+1$, ..., $J$], where $M$ is is the projection quantum number associated with the number of atomic excitations (Fig.~1c). The states with maximum $J$ are symmetric under permutation of atoms, forming the so-called superradiant ladder. Decays between states with the same $J$ (Extended Data Fig.~\ref{decayrates}a) are all collectively enhanced beyond $\Gamma_c$, and in particular, we call such decays within the superradiant ladder superradiance. Meanwhile, any process that does not conserve $J$ is forbidden by symmetry to occur collectively, and must occur via individual dissipation such as spontaneous emission (Extended Data Fig.~\ref{decayrates}b, d, e) or dephasing (Extended Data Fig.~\ref{decayrates}c, f) \citemethods{Zhang2018}. Since the system starts in the ground state and the coherent laser drives the system up the superradiant ladder, the states with $J<N/2$, which form the subradiant subspace, can only be populated through decoherence. In particular, the states $|J, -J\rangle$ in the subradiant subspace cannot collectively decay, and thus are the long-lived dark subradiant states.

Here we also clarify our reasoning for the nomenclature used for the Dicke states. The superradiant ladder is comprised of the states with $J=N/2$, and decays between them are all superradiant. Technically, Dicke defined the $J=N/2, M=0$ state to be the superradiant state \cite{Dicke1954}, however for our purposes we consider all of the enhanced, coherent decays within the ladder to be superradiant as they are enhanced beyond the single-atom decay. Meanwhile, the subradiant subspace is defined as the space formed by the rest of the states, as such states cannot be driven collectively with a coherent drive. We note that decays within the subradiant subspace are not always slower than $\Gamma_c$. In fact, all of the decays within the same $J$ are faster than $\Gamma_c$ even in the subradiant subspace, as shown in Extended Data Fig.~\ref{decayrates}a for $J\leq2$.

Strictly speaking, subradiance is defined as inhibition of emission due to the destructive interference among indistinguishable emitters. By this definition, subradiant decay is forbidden and cannot be observed. However, there are some processes that can break subradiance in order for us to observe that there was suppression of decay. Hence, the experimentally observed slow decay is due to dephasing and individual spontaneous emission processes from the dark subradiant states ($J< N/2, M=-J$). For simplicity, in the main text we refer to this decay as subradiant decay or subradiance, as they provide evidence of subradiance.

\bibliographystylemethods{naturemag}
\bibliographymethods{apssamp}

\begin{acknowledgments}
We thank A. Ruskuc, T. Xie, C.-J. Wu, O. Vendrell, and R. Finkelstein for discussion. \textbf{Funding:} This work was supported by US Department of Energy, Office of Science, National Quantum Information Science Research Centers, Co-design Center for Quantum Advantage (contract number DE-SC0012704), Institute of Quantum Information and Matter, an NSF Physics Frontiers Center (PHY-1733907) with support from the Moore Foundation and by the Office of Naval Research awards no. N00014-19-1-2182 and N00014-22-1-2422, and the Army Research Office MURI program (W911NF2010136). The device nanofabrication was performed in the Kavli Nanoscience Institute at the California Institute of Technology. M.L. acknowledges the support from the Eddleman Graduate Fellowship. R.F. acknowledges the support from the JASSO graduate scholarship. J.R. acknowledges the support from the Natural Sciences and Engineering Research Council of Canada (NSERC) (PGSD3-502844-2017). J.C. acknowledges support from the IQIM postdoctoral fellowship. \textbf{Author Contributions:} A.F. conceived the experiment. M.L. and R.F. built the experimental setup, performed the measurements, and analyzed the data. J.R. fabricated the device. M.L., R.F., B.Z., M.E., J.C., and A.F., interpreted the results. M.L., R.F., J.C., and A.F. wrote the manuscript with inputs from all authors. All work was supervised by J.C. and A.F. \textbf{Competing interests:} The authors declare no competing interests. \textbf{Data and materials availability:} The data that support the findings of this study are available from the corresponding authors upon reasonable request.
\end{acknowledgments}

\clearpage
\onecolumngrid
\section*{Extended Data Figures}
\begin{figure}[H]
  \setcounter{figure}{0}
    \renewcommand{\figurename}{Extended Data FIG.}
    \centering
    \includegraphics[width=\linewidth]{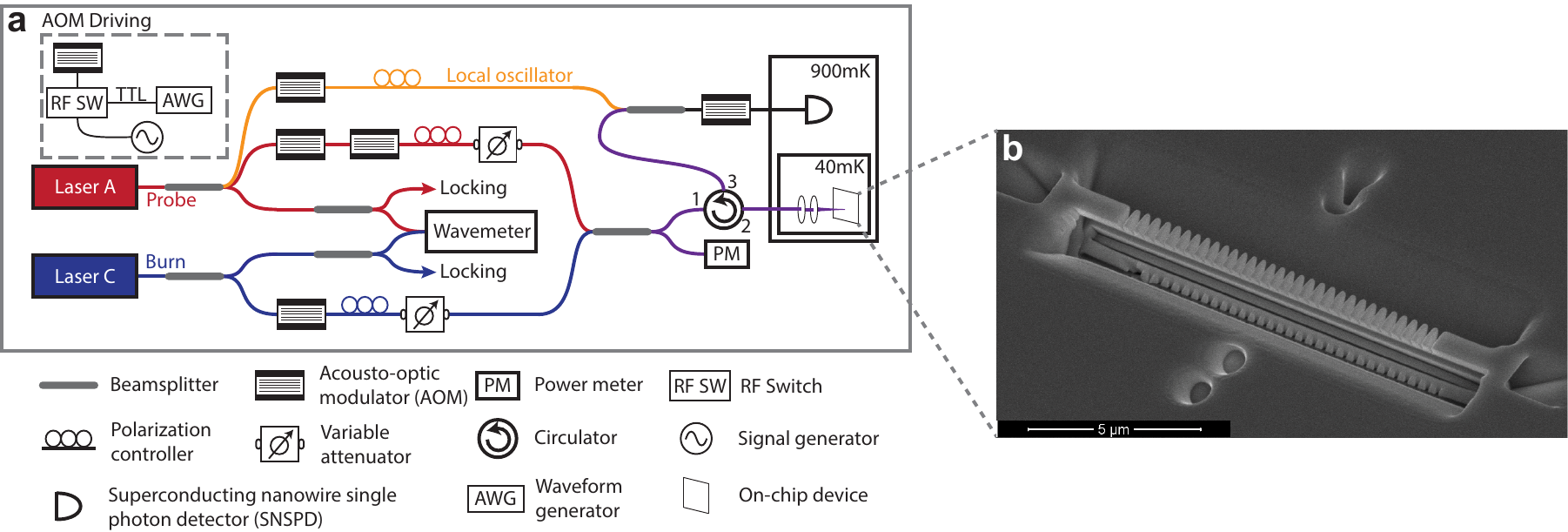}
    \caption{\textbf{Experimental setup.} {\bf a,} Laser A addresses the A transition. Optical pulses are generated using AOMs, which are driven with gated RF sources. Part of it can be split off for use as a local oscillator for heterodyne measurements. A second laser, Laser C can be used to perform optical hole burning on the C transition. The combined light is sent through a circulator, to the device, and the reflected light is sent to a superconducting nanowire single photon detector for time-resolved photon counting. {\bf b,} Scanning electron microscope image of the device.}
    \label{setup}
\end{figure}

\begin{figure}[H]
    \renewcommand{\figurename}{Extended Data FIG.}
    \centering
    \includegraphics[width=0.5\linewidth]{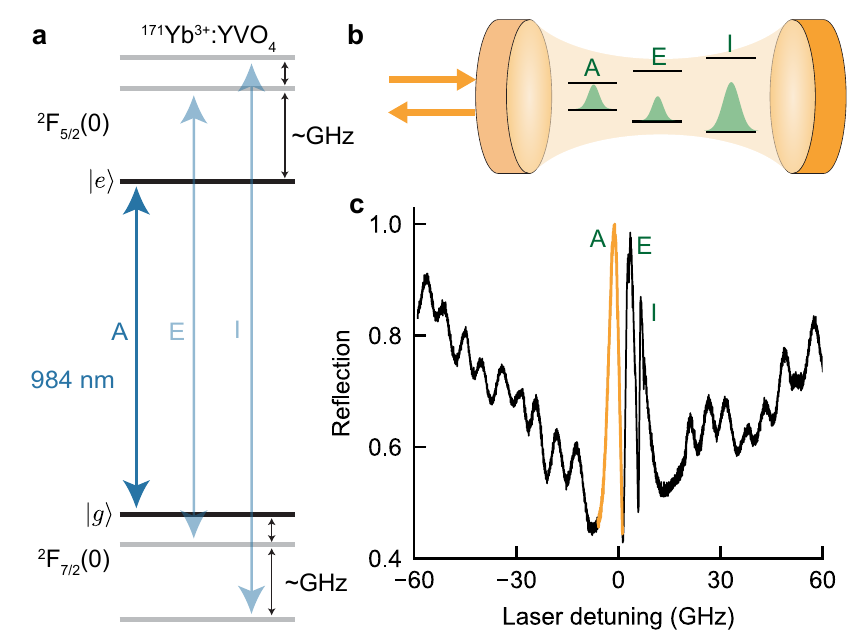}
    \caption{\textbf{Cavity-ion coupling.} {\bf a,} Energy level spectrum of $^{171}$Yb$^{3+}$:YVO$_4$ at zero external magnetic field. The optical transitions whose polarizations are along the cavity mode are shown in blue (A, E, I). Since the three transitions are separated by $\sim$ GHz, larger than the inhomogeneous broadening, each transition is spectrally well-resolved and can be regarded as an isolated, effective two-level system; for example the levels labelled $\ket{e}$ and $\ket{g}$ form a two-level system for the A transition. {\bf b,} Schematic showing the relative population of ions in each transition, where the I transition has double the population due to a doubly-degenerate ground state. {\bf c,} Cavity reflection spectrum at weak laser power reveals three DIR peaks corresponding to the A, E, and I transitions. The peak corresponding to A is marked with orange, as we focus on this transition in the main text.}
    \label{ioncavitycoupling}
\end{figure}

\begin{figure*}
    \centering
    \renewcommand{\figurename}{Extended Data FIG.}
    \includegraphics[width=0.5\linewidth]{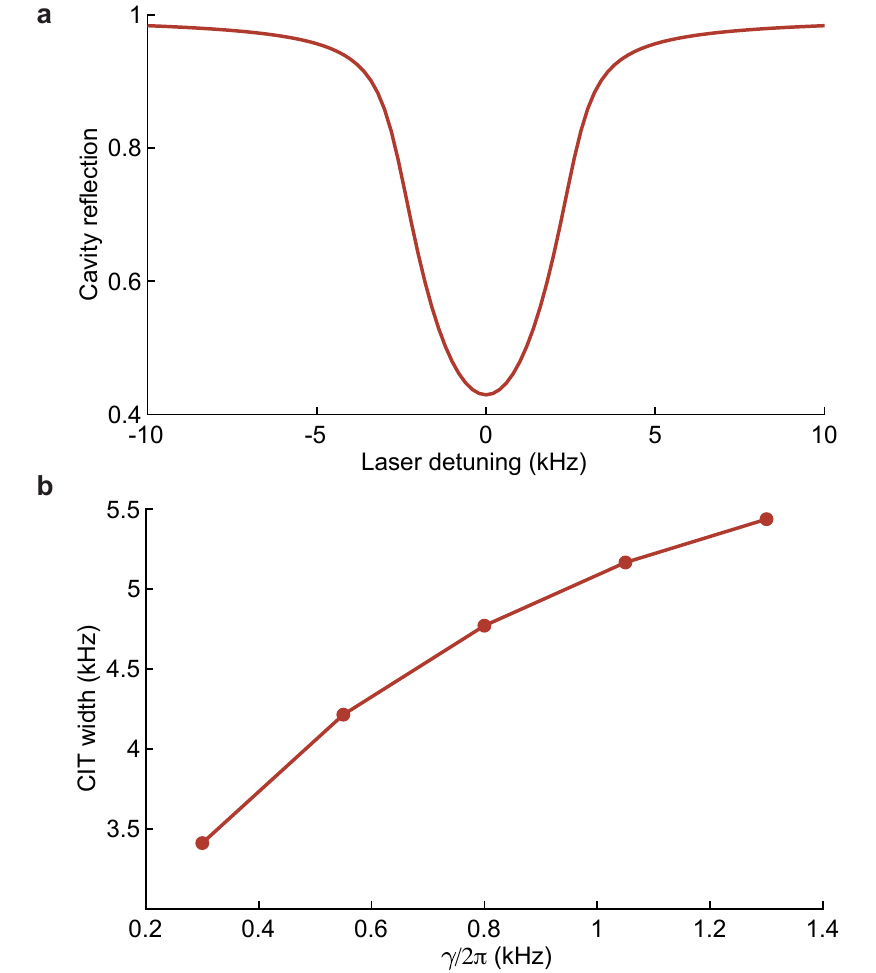}
    \caption{{\bf Numerical simulation under the mean-field approximation of cavity reflection for a system with higher cooperativity.} {\bf a,} Simulated cavity reflection showing narrow CIT for $\Delta_\text{inh}/(2\pi C)=1$ kHz and $\gamma/2\pi=0.3$ kHz. {\bf b,} CIT width with varying $\gamma$ in this high cooperativity system, showing a strong dependence on $\gamma$. See Supplementary Information for simulation details.}
    \label{gamma}
\end{figure*}

\begin{figure*}
    \centering
    \renewcommand{\figurename}{Extended Data FIG.}
    \includegraphics[width=1\linewidth]{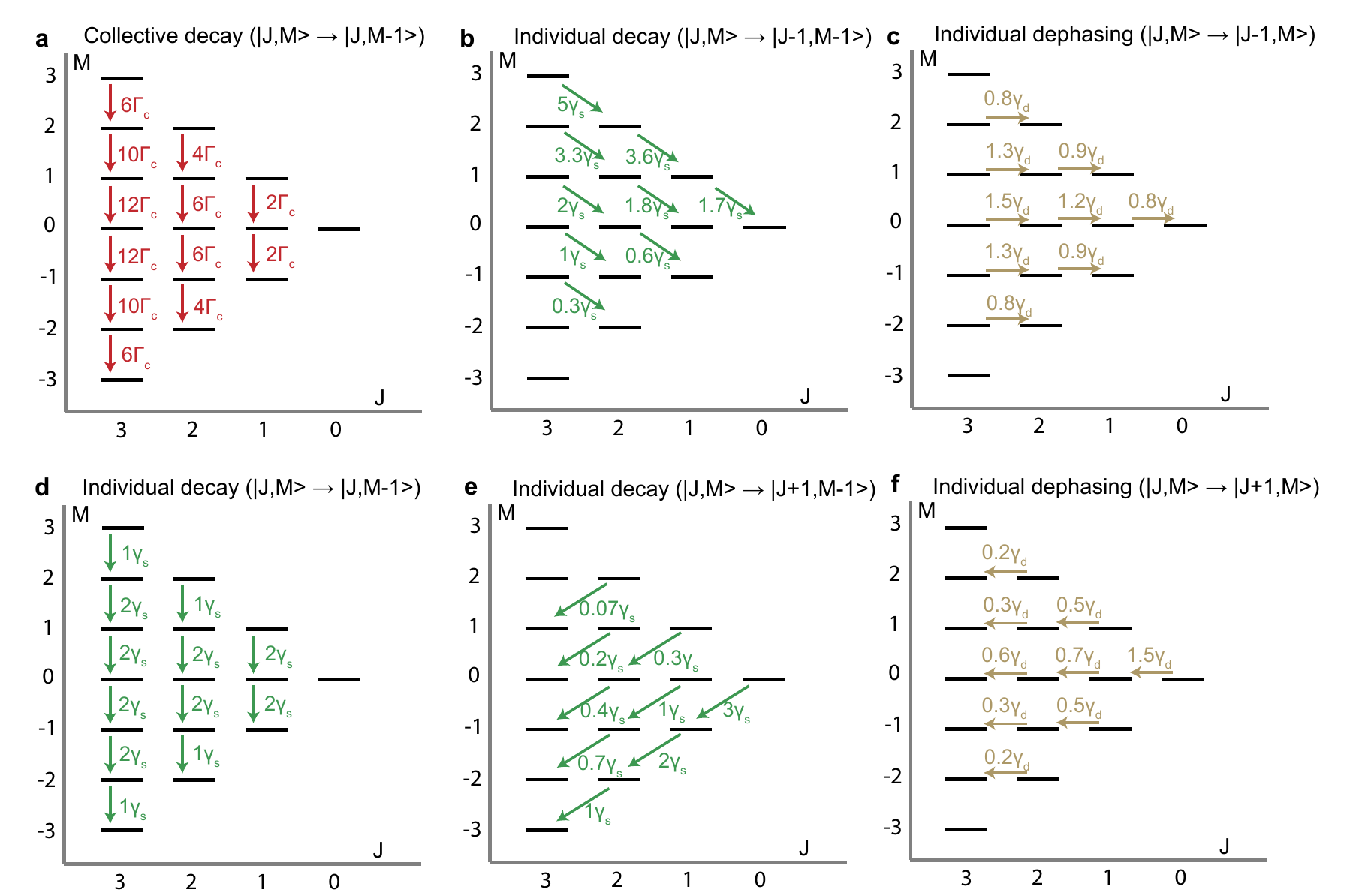}
    \caption{{\bf Decay rates between Dicke states in a bad cavity regime.} Here, Dicke states are formed by 6 identical two-level systems. {\bf a,} Collective decay (red) governed by $\Gamma_c=4g^2/\kappa$ decays vertically, preserving total spin $J$. {\bf b, d, e,} Individual decay (green) governed by spontaneous emission $\gamma_s$ decays diagonally, and {\bf c, f,} individual dephasing (beige) $\gamma_d$ couples neighboring $J$ states with the same $M$. With higher $M$, the diagonal decay rates are faster towards larger $J$, and slower towards smaller $J$.}
    \label{decayrates}
\end{figure*}

\clearpage

\begin{figure*}
    \centering
    \renewcommand{\figurename}{Extended Data FIG.}
    \includegraphics[width=0.5\linewidth]{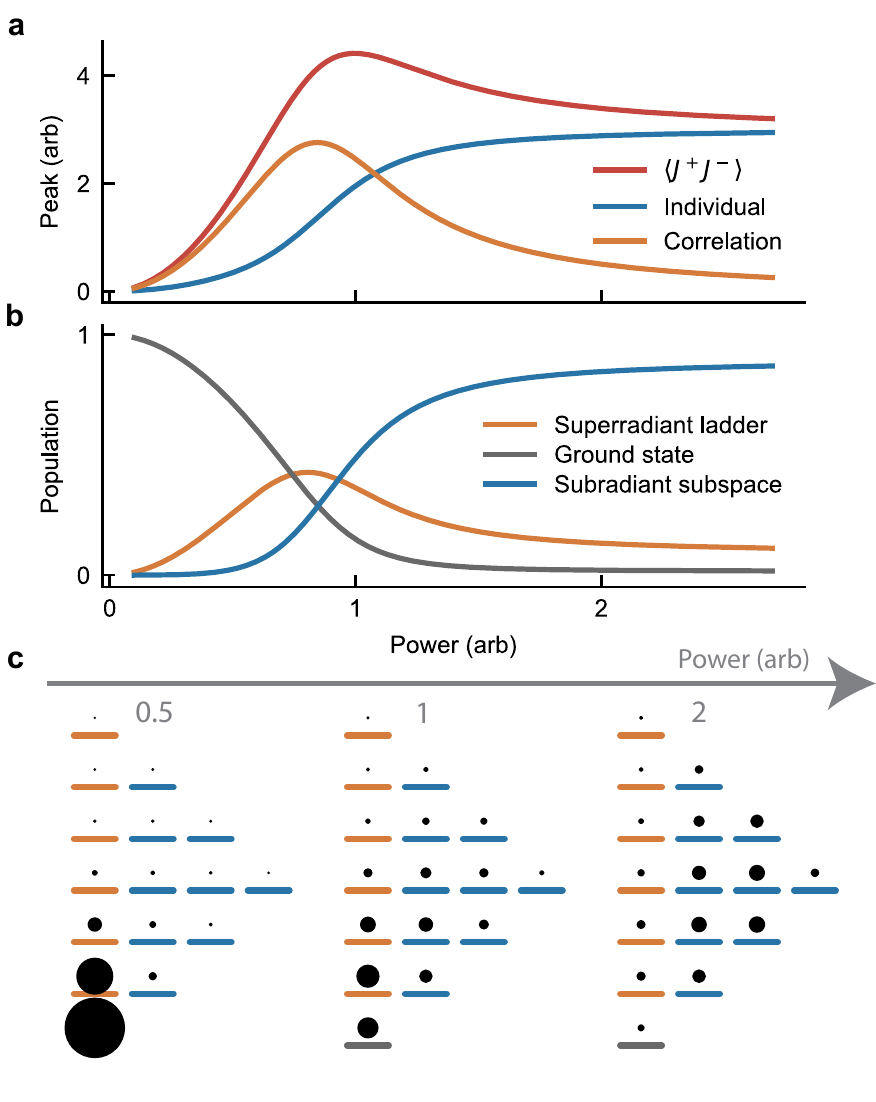}
    \caption{{\bf Master equation simulation of dissipative many-body dynamics.} {\bf a,} Simulation of 6 identical ions excited with a long (50 $\mu$s) pulse. The peak of total squared atomic polarization $\langle J^+J^-\rangle$ (red) is plotted as a function of excitation power along with individual (blue) and correlation (orange) terms from Eq.~\ref{cor}, where the peak amplitudes are calculated from the values right after the excitation pulse, to emulate the peak counts measurements.  {\bf b,} Simulated population dynamics of different subspaces in the Dicke basis as a function of excitation power. Note that the superradiant ladder here refers to all of the states in the $J=3$ manifold except the ground state. We find that the evolution of the superradiant population with power aligns with the correlation term in {\bf a}. {\bf c,} Simulated Dicke state population distribution for different powers. The size of the black circles represents the relative population weights at the end of the excitation pulse. With low power (Power = 0.5 a.u., left), primarily the lower excitation superradiant ladder (orange bars) is populated. With increased power (Power = 1 a.u., middle), the subradiant subspace (blue bars) begins to populate, including the long-lived dark subradiant states. At high power (Power = 2 a.u., right) the system approaches a completely mixed state. Here, the population distribution appears to be unequal among the Dicke states due to the varying degeneracies of the states in the subradiant subspace.}
    \label{sixmaster}
\end{figure*}

\begin{figure*}
    \centering
    \renewcommand{\figurename}{Extended Data FIG.}
    \includegraphics[width=0.5\linewidth]{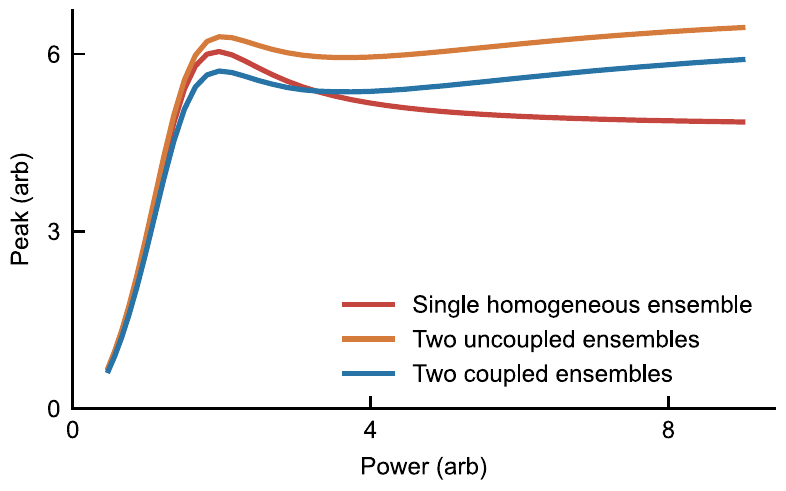}
    \caption{{\bf Comparison of simulated uncoupled and coupled ensembles.} Master equation simulation of the peak counts with power for a single homogeneous ensemble of $5$ ions (red), two {\it coupled} subensembles of $5$ ions at 0 MHz and $2$ ions detuned by $5$ MHz (blue), and two {\it uncoupled} subensembles of $5$ ions at $0$ MHz and $2$ ions detuned by $5$ MHz (orange). Here uncoupled refers to the fact that the peak emission is simulated separately for the subensembles of $5$ and $2$ ions, and later added together. Note that the peak emission reflects the cavity population $\langle a^\dagger a\rangle$ (Methods). The qualitatively similar behavior of the uncoupled and coupled subensemble cases motivates the simulation of an inhomogeneous ensemble via incoherent addition of many uncoupled smaller subensembles in Fig.~3e.}
    \label{incadd}
\end{figure*}

\begin{figure*}
    \centering
    \renewcommand{\figurename}{Extended Data FIG.}
    \includegraphics[width=\linewidth]{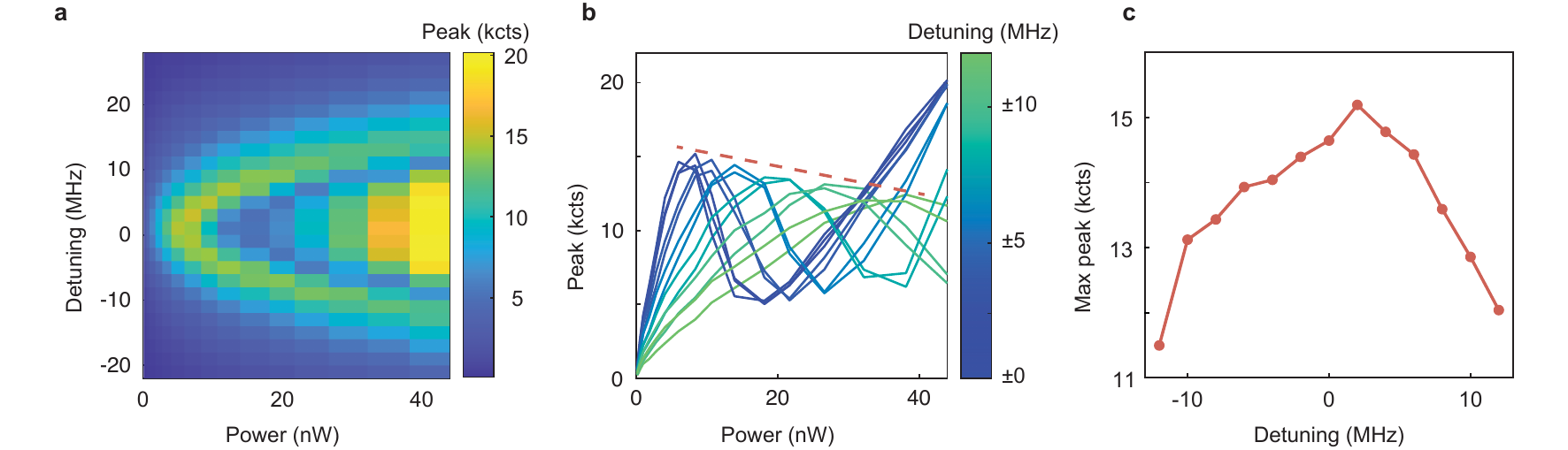}
    \caption{{\bf Frequency and power dependence of peak counts.} \textbf{a,} Peak counts with excitation power and laser detuning. We repeat the pulsed excitation measurement at different frequencies along the inhomogeneous line in the A transition, and observe that the S-curve shifts to higher power. \textbf{b,} Horizontal cuts of \textbf{a} at different detunings. Red dashed line indicates the maximum peak counts for different detunings. \textbf{c,} Extracted local maximum of peak counts as a function of laser detuning. The maximum of peak counts decreases with increased detuning from the center. We note that this behavior differs from Fig.~4b, where while the S-curve also shifted towards higher powers, the maximum of peaks counts \textit{increased} with laser detuning. Here, the S-curve shifts towards higher powers because of the CIT profile; as the laser is detuned from the center, less power enters the cavity, resulting in effectively more power being required to excite the ions. At the same time, the \textit{decrease} of the maximum of peak counts indicates there are less ions resonant with the laser when detuned from the center.}
    \label{vshape}
\end{figure*}
\begin{figure*}
    \centering
    \renewcommand{\figurename}{Extended Data FIG.}
    \includegraphics[width=0.5\linewidth]{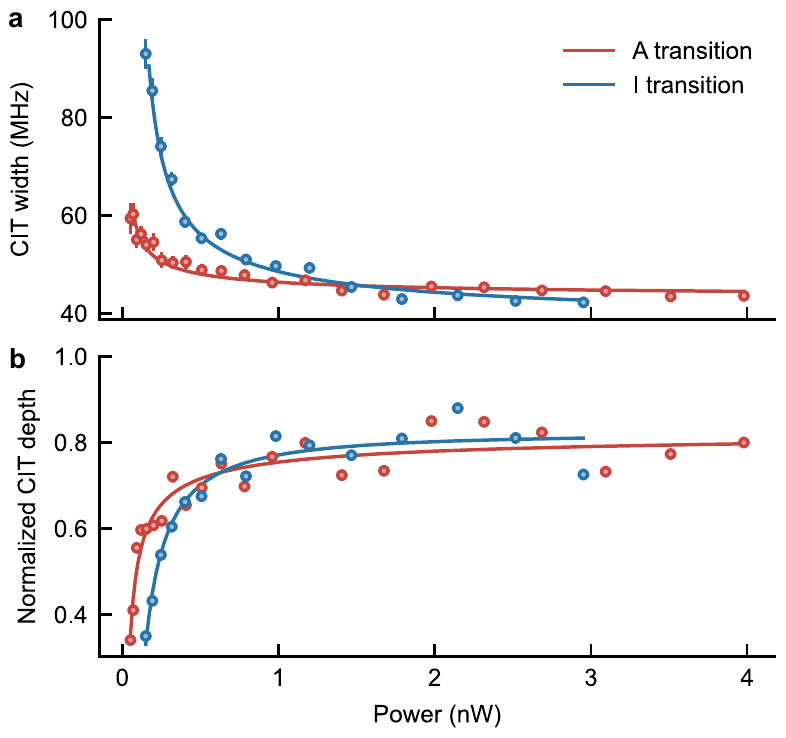}
    \caption{{\bf CIT width and depth comparison between the A and I transitions.} \textbf{a,} Measured CIT width and \textbf{b,} depth with power and corresponding fits for the A (red) and I (blue) transitions. We expect the I transition to have different width and depth values, as the cooperativity is twice as high, but the dephasing rate is expected to be more than $100$ times larger than A. While we find similar depth values, the width differs between the two. In particular, the I transition starts out much broader than A, as expected from the worse coherence properties. Despite this, the minimum width is slightly narrower for the I transition, indicating that indeed the cooperativity is larger for the I transition. Further discussion of the fit parameters are in Methods.}
    \label{AIcomp}
\end{figure*}

\begin{figure}
    \centering
    \renewcommand{\figurename}{Extended Data FIG.}
    \includegraphics[width=0.5\linewidth]{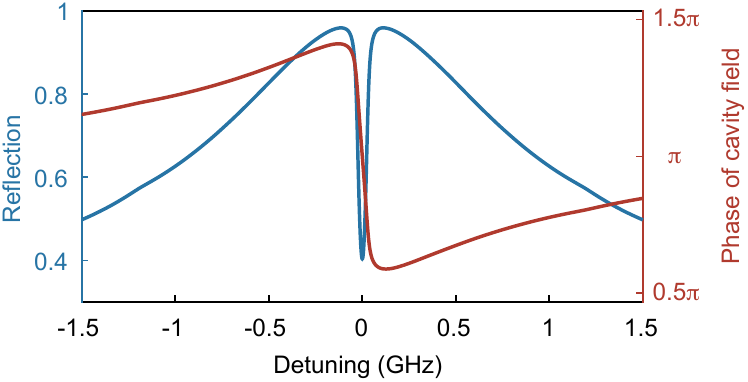}
    \caption{{\bf Numerical simulation of the cavity phase across the CIT.} The simulated cavity phase Arg($a$) (red) across the CIT for a laser power of $0.5$ nW, exhibiting a relative $\pi$ phase shift. The corresponding cavity reflection spectrum (blue) as a function of laser frequency for reference, identical to Fig.~2c purple.}
    \label{phaseshift}
\end{figure}

\begin{figure*}
    \centering
    \renewcommand{\figurename}{Extended Data FIG.}
    \includegraphics[width=\linewidth]{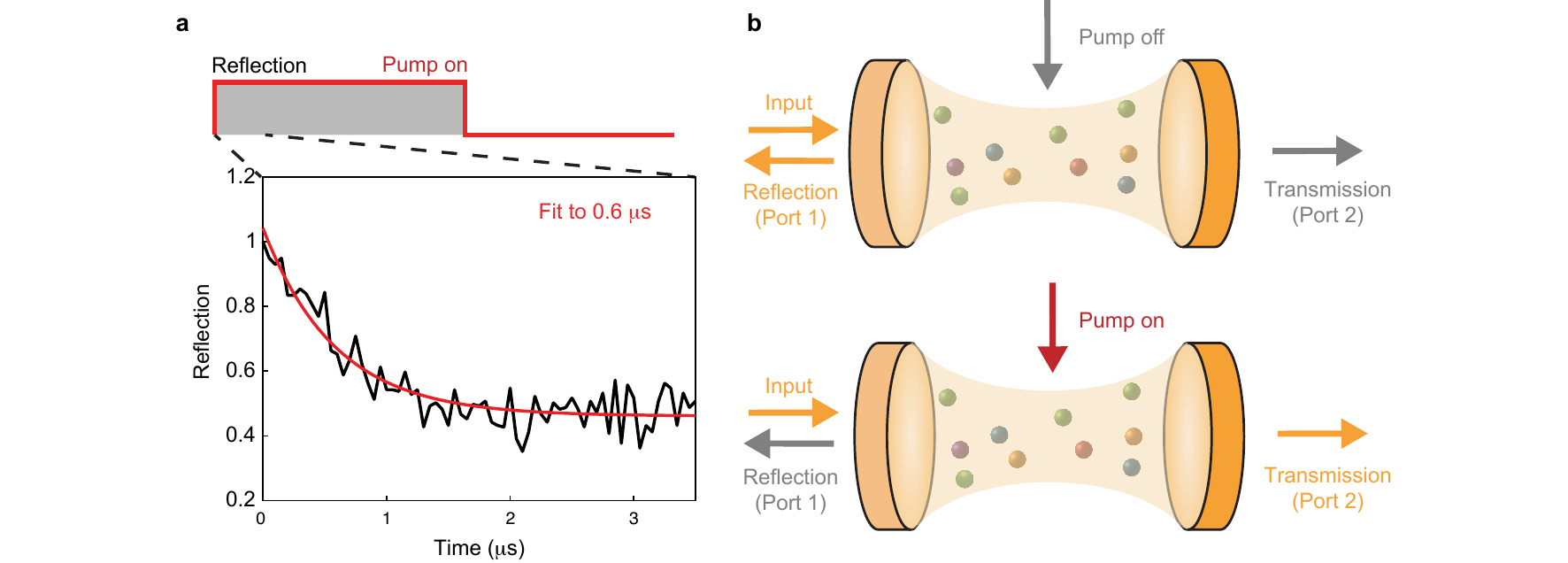}
    \caption{{\bf Optical switch based on CIT.} \textbf{a,} CIT response time. We park the laser at the center of the CIT, and measure the reflection as a function of time after turning the pump on. The reflection signal decreases as the CIT is created in sub-$\mu$s timescales. \textbf{b,} Schematic of an ideal two-port optical switch with an integrated filter with CIT. A transmission port and a transverse pump port are added to realize this application. For the best extinction ratio, the cavity should be two-sided and critically coupled such that $\kappa_1/\kappa=\kappa_2/\kappa=0.5$ where $\kappa_1$ and $\kappa_2$ are the coupling rates from port 1 and 2. The top schematic shows that when the pump is off and DIR is formed, the signal is entirely reflected (port 1). The bottom schematic shows that when the pump is on and CIT is created, the signal is transmitted (port 2) within the CIT window (spectral filter). See Supplementary Information for a more detailed discussion.}
    \label{citresponse}
\end{figure*}

\end{document}


\begin{center}
    {\Large \textbf{Supplementary Information}\par}
\end{center}
\vspace{0.5cm}

\section{System parameters}
The ion-cavity coupling rate $g$ has a particular distribution, arising from inhomogeneity of the cavity field across the device. Simulating the cavity structure in COMSOL, the total YVO$_4$ volume is $3.058\, \text{um}^3$, with total coupling rate $\Omega=2\pi\times9$ GHz, calculated assuming uniformly distributed Yb ions across the YVO$_4$. In order to obtain the distribution of $g$, we generate the histogram of varying $g$ across the entire nanobeam volume of $3.058\, \text{um}^3$. Since there are many ions with very small $g$, we ignore ions with $g<1.39$~MHz which retains $98.4$\% of $\Omega$, resulting in a total coupling rate of $\Omega=2\pi\times8.86$ GHz and the distribution shown in Fig.~S1. The considered ions occupy a physical volume of $0.636\, \text{um}^3$ and their total number is $\approx7\times10^5$. From this, the root mean square of $g$ is estimated to be $\sqrt{\langle g^2 \rangle}=\frac{\Omega}{\sqrt{N}}\approx2\pi\times10.6$ MHz, used later in the estimation of the number of ions participating in superradiance. Additionally, we calculate the optical mode volume to be $V_{\text{mode}}=0.0918\, \text{um}^3$, based on the maximum field strength.
\begin{figure}[h]
    \centering
    \includegraphics[width=\linewidth]{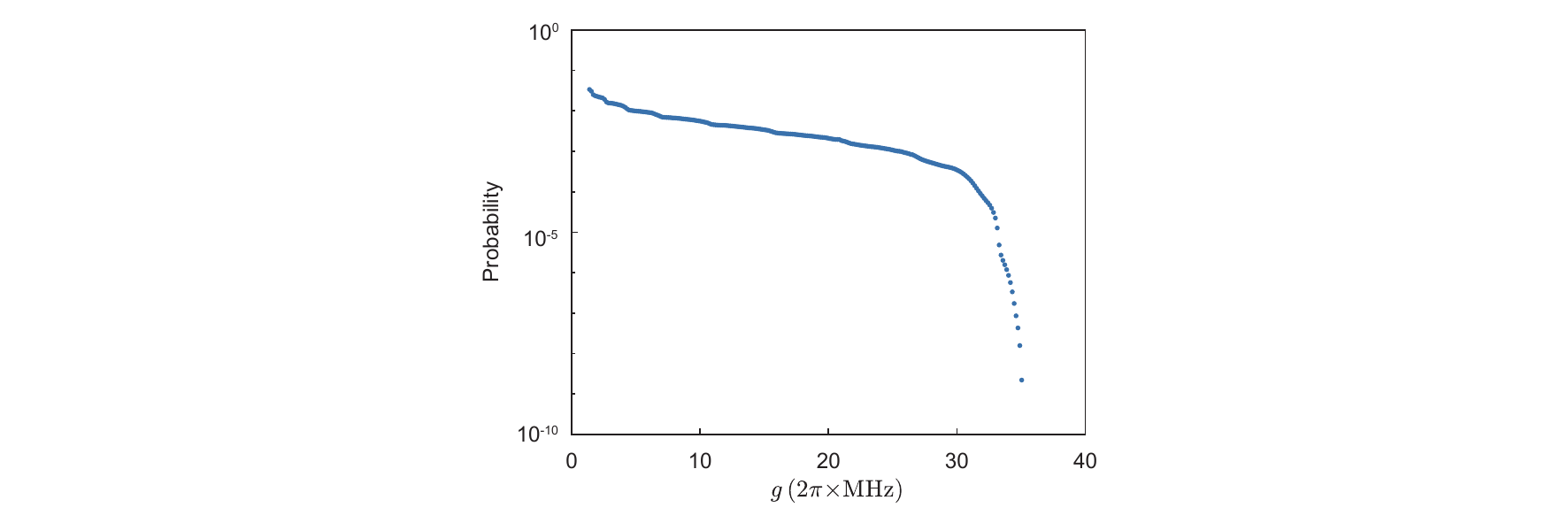}
    \caption{The simulated probability distribution of cavity-ion coupling strength $g$ in the device.}
    \label{g_dist}
\end{figure}

\begin{table}[h]
    \centering
    \begin{tabular}{|p{8cm}|p{2cm}|p{3cm}|}
     \hline
     Parameter Description & Symbol & Value\\
     \hline
     Optical Frequency & $\omega$ & $2\pi\times304500$ GHz\\
     Spontaneous decay rate & $\gamma_s$ & $2\pi\times0.6$ kHz\\
     Excess dephasing rate & $\gamma_d$ & $2\pi\times6$ kHz\\
     Optical inhomogeneous linewidth (FWHM) & $\Delta_{\text{inh}}$ & $2\pi\times150$ MHz\\
     Cavity energy decay rate & $\kappa$ & $2\pi\times44$ GHz\\
     Cavity external coupling rate & $\kappa_c$ & $2\pi\times8.8$ GHz\\
     Maximum Yb ion-cavity coupling rate & $g$ & $2\pi\times35$ MHz\\
     Number of Yb ions in cavity & $N$ & $700000$\\
     Cavity mode volume & $V_\text{mode}$ & $0.0918\mu$m$^3$\\
     Ensemble cooperativity & $C_\text{A}(C_\text{I})$ & $12 (24)$\\
     \hline
    \end{tabular}
    \caption{Relevant system parameters.}
    \label{parameters}
\end{table}

\section{Theoretical Analysis of CIT}
To understand the physical origins of the Collectively Induced Transparency (CIT) dip observed within the broad DIR peak, we start by analyzing simple cases with only a few ions, followed by the large ensemble case with an inhomogeneous distribution.
\subsection{General formalism}
We start with the Tavis-Cummings Hamiltonian for $N$ two-level systems coupled to a single cavity field under the rotating-wave approximation (setting $\hbar=1$):

\begin{equation} \label{h}
    H = \Delta_c a^{\dagger}a + \frac{1}{2}\sum_{j=1}^{N} \Delta_j\sigma_j^z + \sum_{j=1}^{N} g_j (a^{\dagger}\sigma_j^- + \sigma_j^+a)-i\sqrt{\kappa_{c}}A_\text{in}(a^{\dagger}-a).
\end{equation}
In our case, the two-level system consists of the ground ($\ket{g}$) and excited state ($\ket{e}$) of the A transition (see main text). Here $g_j$ is the $j^{\text{th}}$ ion-cavity coupling strength, $a$ is the bosonic cavity field operator, $\sigma_j^{\pm}, \sigma_j^z$ are the $j^{\text{th}}$ spin ladder operators and the Pauli-Z operators, respectively. Note that here we consider the general case where each ion has a different coupling strength $g_j$. $\Delta_c$ is the cavity-laser detuning and $\Delta_j$ is the $j^\text{th}$ ion-laser detuning, accounting for inhomogeneous broadening. $\sqrt{\kappa_c}A_\text{in}$ is the coupling from the input power $P_{\text{in}}$, which is associated with the cavity mean photon number in the absence of ions $\mu$, as:
\begin{equation} \label{power}
   \sqrt{\kappa_c}A_\text{in}=\frac{\kappa}{2}\sqrt{\mu}
\end{equation}
where $\mu$ is defined as
   \begin{equation}
   \mu=\frac{\kappa_c}{(\kappa/2)^2+\Delta_c^2} \frac{P_\text{in}}{\hbar \omega}\approx \frac{\kappa_c}{(\kappa/2)^2} \frac{P_\text{in}}{\hbar \omega}.
\end{equation}
Here, three variables $A_\text{in}$, $P_\text{in}$ and $\mu$ all represent the excitation power, and they are related to each other by system parameters. We will use $\mu$ to represent power in the following theoretical analysis for brevity. 

Starting with Eq.~\ref{h}, we obtain the equations of motion for $a$, $\sigma_j^-$ and $\sigma_j^z$ in the Heisenberg picture, with dissipation terms, given as:
\begin{equation} \label{a}
    \dot{a} =  - (i\Delta_c+\frac{\kappa}{2})a-i\sum_{j=1}^{N} g_j\sigma_j^- - \frac{\kappa}{2}\sqrt{\mu}
\end{equation}
\begin{equation} \label{sigma-}
    \dot{\sigma}_j^- = -(i\Delta_j+\gamma)\sigma_j^- +ig_j\sigma_j^za
\end{equation}
\begin{equation} \label{sigmaz}
    \dot{\sigma}_j^z = 2ig_j(a^{\dagger}\sigma_j^- - \sigma_j^+a)-\gamma_s(1+\sigma_j^z)
\end{equation}
where $\gamma=\frac{\gamma_s}{2}+\gamma_d$. Using Eq.~\ref{power} and input-output formalism, $A_\text{out} = \sqrt{\kappa_c}a + A_\text{in}$, we obtain the cavity reflection:
\begin{equation}
    R=\left|\frac{\langle A_\text{out}\rangle}{\langle A_\text{in}\rangle}\right|^2=\left|\frac{2\kappa_{c}}{\kappa\sqrt{\mu}}\langle a \rangle+1\right|^2
\end{equation}
where cavity field $a$ is determined by solving Eqs.~\ref{a}-\ref{sigmaz} and thus depends on the state of ions. 

\subsection{Weak-excitation case (DIR)}
\begin{figure} 
    \centering
    \includegraphics[width=\linewidth]{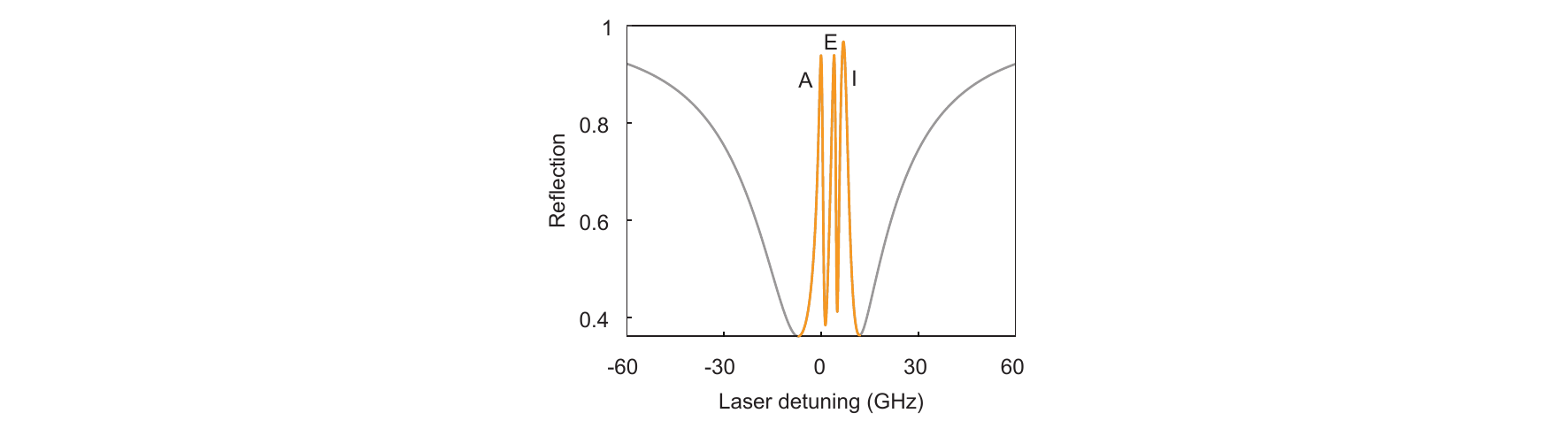}
    \caption{Theoretical cavity reflection as a function of laser detuning in the low excitation regime (DIR), showing ion-cavity coupling. Three DIR peaks (orange) corresponding to the A, E, and I transitions of a Yb ion are identified using Eq.~\ref{refl}, showing high cavity reflectivity.}
   \label{ion-cavity}
\end{figure}

Under the weak-excitation condition ($\langle\sigma_z\rangle\approx-1$), an analytical expression for the cavity reflection $R(\omega)$ as a function of laser frequency $\omega$ is derived in \cite{Diniz2011} as: 
\begin{equation} \label{refl}
    R(\omega)=\left| 1-\frac{i\kappa_c}{\omega-\omega_c+i\kappa/2-W_{\text{A}}(\omega)-W_{\text{E}}(\omega)-W_{\text{I}}(\omega)} \right| ^2
\end{equation}
where $W_\text{X}$ accounts for the coupling of an $\text{X}\;(\text{X}=\text{A},\text{E},\text{I})$ transition to each of the inhomogeneously broadened ensembles:
\begin{equation}\label{Wabsorption}
    W_\text{X}(\omega)=\frac{\Omega_\text{X}^{2}}{\omega-\omega_\text{X}+i\gamma_X+i\Delta_\text{X}/2}.
\end{equation}
Here, $\Delta_X$ is the full-width half-maximum (FWHM) of a Lorentzian distribution, $\Omega_\text{X}=\sqrt{\sum{g_\text{X}^2}}$ is the total ion-cavity coupling rate, $\gamma_\text{X}$ is the dephasing rate, and $\omega_\text{X}$ is the center frequency of an $X$ transition. From COMSOL simulations, we estimate that $\Omega_\text{A}=\Omega_\text{E}=\Omega_\text{I}/\sqrt{2}=2\pi\times4.4$ GHz, where the $1/\sqrt{2}$ factor in $\Omega_\text{I}$ comes from the double degeneracy in that ground state of the I transition. We plot the cavity reflection using Eq.~\ref{refl} in Fig.~\ref{ion-cavity}, which qualitatively matches the experimental data in Extended Data Fig.~2c. From this, the cooperativity $C_\text{X}=\frac{|W_\text{X}(\omega=\omega_0)|}{\kappa/2}$ \cite{Miyazono17} of the three transitions is computed as $C_\text{A,E}\approx 12, C_\text{I}\approx24$.

\subsection{Driven single-ion case}
We would now like to see if CIT appears for just a few ions, starting by considering a single ion coupled to the cavity. Since there is only one ion, the subscript $j$ is dropped. In the fast-cavity limit, we adiabatically eliminate the cavity mode by setting $\dot{a}=0$, and using the fact that $\Delta_c\ll\kappa$ we get
\begin{equation} \label{ae1}
a=\frac{-ig\sigma^--\frac{\kappa}{2}\sqrt{\mu}}{i\Delta_c+\frac{\kappa}{2}}\approx\frac{-ig\sigma^-}{\frac{\kappa}{2}}-\sqrt{\mu}.
\end{equation}
Substituting  Eq.~\ref{ae1} into Eq.~\ref{sigma-} and Eq.~\ref{sigmaz}, we obtain linear differential equations for $\sigma^z$ and $\sigma^-$:
\begin{equation} \label{sigma-1}
    \dot{\sigma}^- = -(i\Delta+\gamma+\frac{2g^2}{\kappa})\sigma^- -ig\sigma^z\sqrt{\mu}
\end{equation}
\begin{equation} \label{sigmaz1}
    \dot{\sigma}^z = -(\gamma_s+\frac{4g^2}{\kappa})(1+\sigma^z)+
    2ig\sqrt{\mu}(\sigma^+-\sigma^-).
\end{equation}
Then we solve the expectation value of the operators in the steady-state $ \langle \dot{\sigma}^z\rangle  = \langle \dot{\sigma}^- \rangle =0$ as:

\begin{equation} 
    \langle \sigma^z \rangle = -\frac{1}{1+\frac{4g^2\mu}{\left(\gamma_s+\frac{4g^2}{\kappa}\right)}\frac{\left(\gamma+\frac{2g^2}{\kappa}\right)}{\left(\Delta^2+\left(\gamma+\frac{2g^2}{\kappa}\right)^2\right)}}
\end{equation}
\begin{equation} 
    \langle \sigma^-\rangle = \frac{ig\sqrt{\mu}}{\left(i\Delta+\gamma+\frac{2g^2}{\kappa}\right)}
    \frac{1}{\left(1+\frac{4g^2\mu}{\left(\gamma_s+\frac{4g^2}{\kappa}\right)}\frac{\left(\gamma+\frac{2g^2}{\kappa}\right)}{\left(\Delta^2+\left(\gamma+\frac{2g^2}{\kappa}\right)^2\right)}\right)}.
\end{equation}
The cavity reflection is
\begin{equation} \label{r1}
    R=\left|\frac{-4i\kappa_c}{\kappa^2\sqrt{\mu}}g\langle \sigma^-\rangle-\frac{\kappa_c}{i\Delta_c+\frac{\kappa}{2}}+1\right|^2
\end{equation}
where $\frac{-4i\kappa_c}{\kappa^2\sqrt{\mu}}g\langle\sigma^-\rangle$ is the contribution from the ion, which is related to the power, and $-\frac{\kappa_c}{i\Delta_c+\frac{\kappa}{2}}+1$ is the contribution from the bare cavity. We plot the cavity reflection for varying $\sqrt{\mu}$ using Eq.~\ref{r1} in Fig.~\ref{ion123}a. We see that in the weak-excitation regime ($\sqrt{\mu}\rightarrow0$, black), there is a DIR centered at the ion frequency, reaching unit cavity reflection. Upon increasing the power, the entire DIR profile decreases, and finally disappears, as the cavity reflection reaches the bare cavity value of $R=\left|-\frac{2\kappa_c}{\kappa}+1\right|^2$ ($R\approx0.36$ for our case). This occurs when we fully saturate the ion.

\subsection{Driven two and three ion cases}
\begin{figure}[t]
    \centering
    \includegraphics[width=\linewidth]{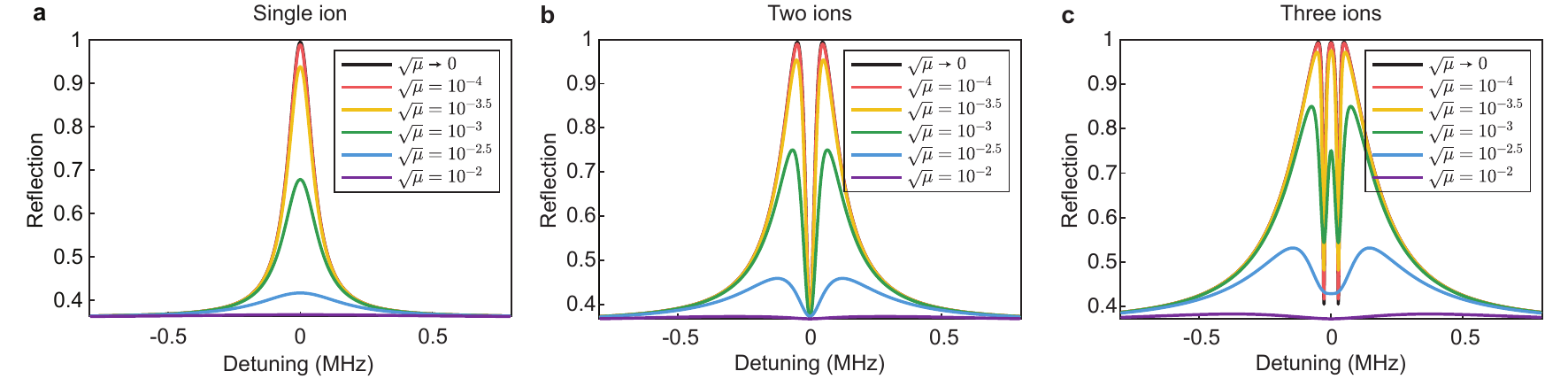}
    \caption{{\bf a,} Cavity reflection spectrum of a single ion coupled to the cavity with rate $g=2\pi\times35$ MHz, using Eq.~\ref{r1}. DIR is observed for small powers, and decreases with increasing powers. {\bf b,} Simulation of the cavity reflection spectrum of two symmetrically detuned ions at $\pm2\pi\times0.048$ MHz. There are two DIR peaks at each of the ion frequencies, and destructive interference between two ions form a dip at $0$ MHz. {\bf c,} Three ion simulation, adding an ion at $0$ MHz. The dip from the two ion case disappears at low powers due to DIR from the added third ion.}
    \label{ion123}
\end{figure}

Since there is no dip for the single ion case, we add another ion, where the ions are now symmetrically detuned around $0$ MHz, with detunings $\pm 2\pi\times0.048$ MHz. The cavity reflection for $\sqrt{\mu}\rightarrow0$ is plotted using an analytical expression under the weak-excitation condition, similar to Eq.~\ref{refl}. Additionally, we sweep the power from $\sqrt{\mu}=10^{-4}$ to $\sqrt{\mu}=10^{-2}$ and simulate using the master equation (Section \ref{Master}). First, we note that $\sqrt{\mu}=10^{-4}$ is very close to the weak-excitation limit. In this case, we observe two DIRs centered around each ion frequency. The two DIRs overlap at zero detuning where the ions destructively interfere, producing a narrow window even with low excitation, indicating that this is not exactly CIT (Fig.~\ref{ion123}b). When $\sqrt{\mu}$ is increased, both DIRs decrease, similar to the single ion case.

We now add a third ion between the two ions at zero detuning, where a third DIR peak appears (Fig.~\ref{ion123}c). At the same time, two sharp dips appear between the first/second ion and second/third ion. When $\sqrt{\mu}$ is increased, all three DIR peaks decrease where the center peak decreases the fastest. This eventually leads to a single, wide dip at $\sqrt{\mu}=10^{-2.5}$. From the results of the two and three ion cases, we see hints of the origins of CIT, from the destructive interference between different ions.
\subsection{Driven multi-ion case}
\subsubsection{Analytical derivation} \label{AD}
From the simple cases analyzed above, we get the intuition that an inhomogeneously broadened ensemble of ions is required for a narrow dip to form. To see this more rigorously, we derive the analytical expression for CIT under certain conditions. We first set the steady state of Eqs.~\ref{a} -- \ref{sigmaz} ($\dot{a}=\dot{\sigma}^z=\dot{\sigma}^-=0$), which gives $2N+1$ equations for $2N+1$ variables:
\begin{equation} \label{a0}
     - (i\Delta_c+\frac{\kappa}{2})\langle a \rangle -i\sum_{j=1}^{N} g_j\langle \sigma_j^-\rangle -\frac{\kappa}{2}\sqrt{\mu} = 0
\end{equation}
\begin{equation} \label{sigma-0}
    -(i\Delta_j+\gamma)\langle \sigma_j^- \rangle +ig_j\langle \sigma_j^z \rangle \langle a\rangle = 0
\end{equation}
\begin{equation} \label{sigmaz0}
     2ig_j(\langle a^{\dagger}\rangle \langle \sigma_j^-\rangle - \langle \sigma_j^+\rangle \langle a\rangle)-\gamma_s(1+\langle \sigma_j^z\rangle) =0.
\end{equation}
Note that here we adopt the hypothesis that quantum correlation between atomic and cavity field operators can be neglected ($\langle\sigma_j^z  a\rangle=\langle\sigma_j^z \rangle \langle a\rangle$ and $\langle a^{\dagger} \sigma_j^-\rangle = \langle a^{\dagger}\rangle \langle \sigma_j^-\rangle $) \cite{allen1987}. 
Eliminating $\langle \sigma_j^-\rangle$, we get the equation relating $\langle \sigma_j^z\rangle$ and $\langle a\rangle$ as:
\begin{equation} \label{az}
     \langle a\rangle =\frac{-\sqrt{\mu}}{1-\sum_{j=1}^{N} \frac{2g_j^2\langle \sigma_j^z\rangle}{\kappa(\gamma+i\Delta_j)}}
\end{equation}
\begin{equation} \label{za}
     \langle \sigma_j^z\rangle=-\frac{1}{1+\frac{4g_j^2\langle a^{\dagger}\rangle \langle a\rangle}{\gamma_s}\left(\frac{\gamma}{\gamma^2+\Delta_j^2}\right)}.
\end{equation}
We define $\langle a\rangle=-\frac{\sqrt{\mu}}{1+x}$, where
\begin{equation} \label{x}
     x=-\sum_{j=1}^{N} \frac{2g_j^2\langle \sigma_j^z\rangle}{\kappa(\gamma+i\Delta_j)}=\frac{\sum_{j=1}^{N}2ig_j\langle \sigma_j^-\rangle}{\kappa \langle a\rangle}
\end{equation}
which represents the ions' response to the cavity field. Here,  we also ignore $\Delta_c$ for simplicity based on the fact that $\Delta_c \ll \kappa$. We can get an implicit equation for $x$ as:
\begin{equation}
     x=\sum_{j=1}^{N} \frac{2g_j^2}{\kappa(\gamma+i\Delta_j)}\frac{1}{\left(1+\frac{4g_j^2\mu}{|1+x|^2\gamma_s}\left(\frac{\gamma}{\gamma^2+\Delta_j^2}\right)\right)}
\end{equation}
where the distribution of coupling strength $g_j$ is not correlated with the distribution of ion frequency detuning $\Delta_j$, as the physical position of the ion in the cavity is unrelated to its resonance frequency. We first consider $g_j=g$ and define $y(x)=\frac{4g^2\mu\gamma}{|1+x|^2\gamma_s}$ for simplicity, and convert the summation to integral in the limit of large $N$:
\begin{equation} \label{xeq}
     x=\frac{2Ng^2}{\kappa}\int \frac{\rho(\omega)d\omega}{\left(\gamma+i(\omega-\omega_L)\right)\left(1+\frac{y(x)}{\gamma^2+(\omega-\omega_L)^2}\right)}
\end{equation}
where $\omega_L$ is laser frequency and $\rho(\omega)$ is the probability distribution of a given ion frequency $\omega$. Given $x$, the cavity reflection is now simply
\begin{equation} \label{Rx}
     R=\left|-\frac{2\kappa_c}{\kappa}\frac{1}{(1+x)}+1\right|^2.
\end{equation}

We consider a Lorentzian distribution of ions with $\rho(\omega)=\frac{\Delta_{\text{inh}}/2}{\pi}\frac{1}{(\Delta_{\text{inh}}/2)^2+(\omega-\omega_0)^2}$ where $\Delta_{\text{inh}}$ is the FWHM and $\omega_0$ is the center frequency of the ion distribution, which gives us:
    \begin{equation}
     x=\frac{N\Delta_{\text{inh}} g^2}{\pi\kappa}\int_{-\infty}^{+\infty} \frac{ \left( \gamma-i(\omega-\omega_L)\right) d\omega}{\left(\gamma^2+(\omega-\omega_L)^2+y(x)\right)\left((\Delta_{\text{inh}}/2)^2+(\omega-\omega_0)^2\right)}.  
     \end{equation}
Making the approximations $\Delta_{\text{inh}}\gg|\omega_L-\omega_0|,\gamma,\sqrt{y(x)}$ and $y(x)\gg \gamma^2$ as well as $\sqrt{\mu}>\frac{2Ng\sqrt{\gamma_s\gamma}}{\kappa\Delta_{\text{inh}}}$ (summarized and justified below), allows us to solve for $x$ explicitly:
\begin{equation} \label{xlor}
 x=\frac{1}{\frac{\Delta_{\text{inh}}\kappa}{2Ng}\sqrt{\frac{\mu}{\gamma_s\gamma}}-1}+\frac{8i(\omega_L-\omega_0)}{\frac{\Delta_{\text{inh}}^2\kappa}{Ng^2}}.
\end{equation}
Plugging Eq.~\ref{xlor} into Eq.~\ref{Rx}, we get the explicit expression of the reflectivity relative to the laser frequency $\omega_L$ as:
\begin{equation} \label{Rlor}
   R=1+\frac{\frac{\kappa_c}{\kappa}(\frac{\Delta_{\text{inh}}}{C})^2\left(\frac{\kappa_c}{\kappa}-\frac{C\Delta_{\text{CIT}}}{\Delta_{\text{inh}}}\right)}{\left(\frac{\Delta_{\text{CIT}}}{2}\right)^2+(\omega_L-\omega_0)^2}
\end{equation}
where $C=\frac{|W(\omega=\omega_0)|}{\kappa/2}(=\frac{4Ng^2}{\kappa\Delta_{\text{inh}}}$ when $
\Delta_{\text{inh}}\gg\gamma)$ and
\begin{equation} \label{eqwidth}
     \Delta_{\text{CIT}}=\frac{\Delta_{\text{inh}}}{C}\frac{1}{\left(1-C\sqrt{\frac{\gamma_s\gamma}{4g^2\mu}}\right)}.
\end{equation}

According to Eq.~\ref{Rlor}, it is apparent that the profile of the CIT dip is a Lorentzian function with width $\Delta_{\text{CIT}}$ and depth $\eta_{\text{CIT}}$, which is determined by the reflectivity at $\omega_L=\omega_0$ as
\begin{equation} \label{eqdepth}
\begin{split}
    \eta_{\text{CIT}}&=\frac{A}{\left(\frac{\Delta_{\text{CIT}}}{2}\right)^2\eta_\text{bare}}\\
    &=\frac{1}{(1-\frac{\kappa_c}{\kappa})}\left(1-C\sqrt{\frac{\gamma_s\gamma}{4g^2\mu}}-\frac{\kappa_c}{\kappa}\left(1-C\sqrt{\frac{\gamma_s\gamma}{4g^2\mu}}\right)^2\right)
\end{split}
\end{equation}
where the depth has been normalized against the bare cavity depth $\eta_\text{bare}=1-\left(1-\frac{2\kappa_c}{\kappa}\right)^2$.
The lower bound of $\Delta_{\text{CIT}}$ is $\Delta_{\text{CIT, min}}=\frac{\Delta_{\text{inh}}}{C}$, as the laser power $\mu$ becomes large. Additionally in the same limit, the depth reaches $1$.
Here we summarize the assumptions made to derive the analytical expressions of the CIT width and depth.
\begin{subequations} \label{AS}
\begin{align}
  C&\gg1 &\text{High cooperativity}\\
    \left(\frac{\Delta_\text{inh}}{4g}\right)^2\frac{\gamma_s}{\gamma}\gg &\mu \gg \frac{\gamma\gamma_s}{4g^2} &\text{Intermediate power} \\
   \frac{\Delta_\text{inh}}{\gamma}&\gg C &\text{Appreciable inhomogeneity and good coherence}
\end{align}
\end{subequations}
We note that for a uniform (rectangular) ensemble distribution, the CIT depth and width have the same power and cooperativity dependence up to a factor of $\frac{\pi}{2}$. From this we anticipate CIT to be a ubiquitous phenomenon for any distribution satisfying Eq.~\ref{AS}. 

\subsubsection{Inhomogeneity of $g$ distribution}
For simplicity, we have so far assumed that all ions are coupled equally to the cavity ($g_j=g$). Now we consider inhomogeneously distributed $g$, with a probability distribution given by $p(g)$. Using the fact that $p(g)$ is uncorrelated with $\rho(\omega)$, Eq.~\ref{xeq} can be written as:
\begin{equation} \label{xeqg}
     x=\frac{2}{\kappa} \int Ng^2p(g)dg
     \int \frac{\rho(\omega)d\omega}{\left(\gamma+i(\omega-\omega_L)\right)\left(1+\frac{y(x,g)}{\gamma^2+(\omega-\omega_L)^2}\right)}.
\end{equation}
Note that $y(x,g)$ is now also a function of $g$. We can first calculate the integral over frequency $
\omega$, then over $g$ to get the solution of $x$ similar to Eq.~\ref{xlor}:
\begin{equation}
    x=\frac{1}{\frac{\Delta_{\text{inh}}\kappa}{2Ng_\text{avg}}\sqrt{\frac{\mu}{\gamma_s\gamma}}-1}+\frac{8i\omega_L}{\frac{\Delta_{\text{inh}}^2\kappa}{\Omega^2}}
    \end{equation}
    where $\Omega^2=\int Ng^2p(g)dg$, and we define $g_\text{avg}=\int gp(g)dg$.
Now, we obtain the width 
\begin{equation} 
     \Delta_{\text{CIT}}=\Delta_{\text{inh}}\frac{\frac{\Delta_{\text{inh}}\kappa}{4\Omega^2}}{1-\frac{2Ng_\text{avg}}{\Delta_{\text{inh}}\kappa}\sqrt{\frac{\gamma_s\gamma}{\mu}}}.
\end{equation}
From this, we can see that the CIT profile and the minimum width do not depend on the specific distribution of cavity-ion coupling strength, but only the total coupling strength. Ensemble cooperativity $C\propto\frac{\Omega^2}{\kappa\Delta_{\text{inh}}}$ also only depends on the total coupling strength $\Omega^2$, so the expression of the minimum FWHM relative to the cooperativity will still be $\Delta_{\text{CIT, min}}\sim\frac{\Delta_\text{inh}}{C}$, regardless of the specific distribution of $g$.
\subsubsection{Intuitive understanding}
From the above derivations, we see that for high cooperativity ($C>1$), we will get a CIT dip with width approximately equal to the inhomogeneous linewidth divided by $C$. Looking back at Eq.~\ref{x}, the contribution of each ion to $x$ can be extracted using Eq.~\ref{sigma-0} (still assuming $g_j=g$ since $g_j$ and $\Delta_j$ are uncorrelated):

\begin{equation} \label{singphase}
     \frac{\langle\sigma^-_{
     \scalebox{.9}{$\scriptscriptstyle \Delta_j$}
     }\rangle}{\langle a\rangle}=\frac{ig\sigma_{\Delta_j}^z(\gamma-i\Delta_j)}{\gamma^2+\Delta_j^2}.
\end{equation}
Here, we use notation $\langle \sigma_j^-\rangle=\langle \sigma_{
\scalebox{.9}{$\scriptscriptstyle \Delta_j$}
}^-\rangle$ and $\langle \sigma_j^z\rangle =\langle \sigma_{
\scalebox{.9}{$\scriptscriptstyle \Delta_j$}
}^z\rangle$, where the subscript now denotes the ion-laser detuning. For ions with detuning much larger than a homogeneous linewidth $|\Delta_j|\gg\gamma$, if a ion is blue-detuned ($\Delta_j>0$), its atomic coherence phase $\text{arg}\Big(\frac{\langle\sigma^-_{
\scalebox{.9}{$\scriptscriptstyle \Delta_j$}
}\rangle}{\langle a\rangle}\Big)\to-\pi$ while its red-detuned pair has $\text{arg}\Big(\frac{\langle \sigma^-_{
\scalebox{.9}{$\scriptscriptstyle -\Delta_j$}
}\rangle}{\langle a\rangle}\Big)\to 0$. Their amplitudes are $\Big|\frac{\langle \sigma^-_{
\scalebox{.9}{$\scriptscriptstyle \Delta_j$}
}\rangle }{\langle a\rangle}\Big|=\Big|\frac{\langle \sigma^-_{
\scalebox{.9}{$\scriptscriptstyle -\Delta_j$}
}\rangle }{\langle a\rangle }\Big| \sim \frac{g}{|\Delta_j|}|\langle \sigma_{
\scalebox{.9}{$\scriptscriptstyle |\Delta_j|$}
}^z\rangle|$. We can see that for the above pair of ions with $\pm\Delta_j$, the phase difference will be $\pi$, such that their phases will cancel and their contribution to the cavity field will be
\begin{equation} \label{pairs}
     \frac{\langle \sigma^-_{
     \scalebox{.9}{$\scriptscriptstyle \Delta_j$}
     }\rangle}{\langle a\rangle}+\frac{\langle \sigma^-_{
     \scalebox{.9}{$\scriptscriptstyle -\Delta_j$}
     }\rangle}{\langle a\rangle}=\frac{ig\langle \sigma_{
     \scalebox{.9}{$\scriptscriptstyle |\Delta_j|$}
     }^z\rangle (2\gamma)}{\gamma^2+\Delta_j^2}\sim\frac{2g\gamma}{\Delta_j^2}
     |\langle \sigma_{\scalebox{.9}{$\scriptscriptstyle |\Delta_j|$}}^z\rangle|
\end{equation}
Note that the phase we talk about here is the phase of $\langle \sigma_j^- \rangle$ relative to $\langle a \rangle$. For the Fig.~2a in the main text, we add a global phase of $\frac{\pi}{2}$ to the ion phases for visual clarity.

To summarize, 
\begin{enumerate}
    \item The contribution from any ion to cavity reflection will decrease as the ion gets excited ($\langle\sigma^z_j\rangle>-1$), and finally disappear when the ion is saturated to the completely mixed state where $\langle \sigma^z_j\rangle=\langle \sigma^-_j\rangle = 0$.
    \item The phase of each ion relative to the cavity field is determined by the detuning of an ion relative to the laser frequency. For blue-detuned ions ($\Delta_j>0$), the phases are between $-\frac{\pi}{2}$ to $-\pi$, while for the red-detuned ions ($\Delta_j<0$), the phases are between $-\frac{\pi}{2}$ to $0$.
    \item A pair of symmetrically detuned ions relative to the laser frequency will have a reduced contribution to the cavity field compared to a single ion due to the phase cancellation, by a factor of $\sim\frac{\gamma}{\Delta_j}$. This is obtained by comparing Eq.~\ref{pairs} to Eq.~\ref{singphase} in the case of $|\Delta_j|\gg \gamma$.
\end{enumerate}
\subsubsection{Mean-field approximation}
In the above derivation of CIT, we have used the mean-field approximation \cite{Glicenstein2020,Santo2020} where the mean values of the products of the operators $a$ and $\sigma_j^z$ are factorized: $\langle \sigma_j^z a\rangle=\langle \sigma_j^z\rangle \langle a\rangle$. In the following, we will justify why such an approximation can be made for this system when considering CIT. To start, we introduce the term $\langle \sigma_j^z a\rangle-\langle \sigma_j^z\rangle \langle a\rangle$ which considers the correlations between $a$ and $\sigma_j^z$, and we would like to show that this term is negligible.
Using adiabatic elimination $a\approx\frac{-2i}{\kappa}\sum_{k=1}^Ng_k\sigma_k^--\sqrt{\mu}$, we expand this as:
\begin{align}
    \langle \sigma_j^z a\rangle-\langle \sigma_j^z\rangle \langle a\rangle &=-\frac{2i}{\kappa}\sum_{k=1}^Ng_k\left( \langle \sigma_j^z \sigma_k^-\rangle-\langle \sigma_j^z\rangle \langle \sigma_k^-\rangle\right) \nonumber \\
    &=-\frac{2i}{\kappa}\left(g_j\left( \langle \sigma_j^z \sigma_j^-\rangle-\langle \sigma_j^z\rangle \langle \sigma_j^-\rangle\right)+\sum_{k\neq j}^Ng_k\left( \langle \sigma_j^z \sigma_k^-\rangle-\langle \sigma_j^z\rangle \langle \sigma_k^-\rangle\right)\right)
\end{align}
The first term is the self-correlation between $\sigma_j^z$ and $\sigma_j^-$ of ion $j$, and the second term is the cross-correlation between different ions. Setting the second term $\langle \sigma_j^z \sigma_k^-\rangle-\langle \sigma_j^z\rangle \langle \sigma_k^-\rangle=0$ ($k\neq j$) indicates that the final states of different ions are not entangled.
For the first term (self-correlation), we use the commutation rule between $\sigma_j^z$ and $\sigma_j^-$ ($\sigma_j^z\sigma_j^-=-\sigma_j^-$) to rewrite it as
\begin{equation}
    -\frac{2ig_j}{\kappa}\left( \langle \sigma_j^z \sigma_j^-\rangle-\langle \sigma_j^z\rangle \langle \sigma_j^-\rangle\right)= \frac{2ig_j}{\kappa}\left( 1+ \langle \sigma_j^z\rangle\right)\langle \sigma_j^-\rangle
\end{equation}
It is true that this term can be nonzero based on the solution in CIT. However, we now want to show that the influence of this term of the final cavity field $\langle a \rangle$ is very small. To do this, we introduce $\delta(\langle \sigma_j^- \rangle)$, which is the modification to $\langle \sigma_j^- \rangle$ if we were to consider this self-correlation term. Using Eq.~\ref{sigma-0},
\begin{equation}
    \delta(\langle \sigma_j^- \rangle) = -\frac{2g_j^2}{(i\Delta_j+\gamma)\kappa}\left(1+\langle \sigma_j^z\rangle \right)\langle \sigma_j^-\rangle
\end{equation}
We can then plug this back into our expression for the cavity field $a$ using Eq.~\ref{a0}, and get the modification to the cavity field introduced by this self-correlation:
\begin{align} \label{da}
    \delta(\langle a\rangle) &= -\frac{2i}{\kappa}\sum_{k=1}^Ng_k\delta(\langle \sigma_k^-\rangle \nonumber \\
    &=\frac{4i}{\kappa^2}\sum_{k=1}^N\frac{g_k^3}{i\Delta_k+\gamma}\left(1+\langle \sigma_k^z \rangle\right)\langle \sigma_k^- \rangle \nonumber \\
    &=-\frac{4}{\kappa^2}\sum_{k=1}^N\frac{g_k^4}{(i\Delta_k+\gamma)^2}\left(1+\langle \sigma_k^z \rangle\right)\langle \sigma_k^z \rangle \langle a \rangle
\end{align}
If we plug solution Eq.~\ref{za} into Eq.~\ref{da}, we get
\begin{equation}
    \delta(\langle a\rangle)=\frac{4}{\kappa^2}\sum_{k=1}^N\frac{g_k^4}{(i\Delta_k+\gamma)^2}\frac{\frac{4g_k^2|
   \langle a \rangle|^2\gamma}{\gamma_s(\gamma^2+\Delta_k^2)}}{\left(1+\frac{4g_k^2|
   \langle a \rangle|^2\gamma}{\gamma_s(\gamma^2+\Delta_k^2)}\right)^2}\langle a \rangle
\end{equation}
Assuming $g_k=g$ for all the ions, we can obtain the relative change of $\langle a \rangle$ is
\begin{equation}
    \left|\frac{\delta(\langle a\rangle)}{\langle a\rangle}\right|<\frac{16g^6\gamma}{\kappa^2\gamma_s}\sum_{k=1}^N\frac{|\langle a\rangle|^2}{(i\Delta_k+\gamma)^2(\gamma^2+\Delta_k^2)}\sim\frac{16Ng^6\gamma}{\kappa^2\gamma_s\Delta_\text{inh}^4}|\langle a\rangle|^2\sim\frac{16Ng^6\gamma}{\kappa^2\gamma_s\Delta_\text{inh}^4}\mu
\end{equation}
Using Eq.~\ref{AS}b where $\mu\ll\left(\frac{\Delta_\text{inh}}{4g}\right)^2\frac{\gamma_s}{\gamma}$ we can then get $ \left|\frac{\delta(\langle a\rangle)}{\langle a\rangle}\right|\ll\frac{Cg^2}{4\kappa\Delta_\text{inh}}\ll 1$, which is validated both in experiment and simulation for our system.
The conclusion is that $\left|\frac{\delta(\langle a\rangle)}{\langle a\rangle}\right| \ll 1$, or that the fractional change in the cavity field due to the self-correlation term will be very small.

\subsubsection{Description of CIT in the good-cavity regime}
We have so far focused on the resonant bad-cavity regime, such that $\Delta_c=\omega_c-\omega_L$, the cavity detuning from the laser, was ignored. Here we generalize the CIT derivation to the case where $\Delta_c$ is considered, and \ref{xeq} becomes:
\begin{equation} \label{xeqnew}
     x=\frac{2Ng^2}{\kappa+2i\Delta_c}\int \frac{\rho(\omega)d\omega}{\left(\gamma+i(\omega-\omega_L)\right)\left(1+\frac{y(x)}{\gamma^2+(\omega-\omega_L)^2}\right)}
\end{equation}
where now $y(x)=\frac{4g^2\mu\gamma}{|1+x|^2\gamma_s(1+4\Delta_c^2/\kappa^2)}\approx\frac{4g^2\mu\gamma}{|1+x|^2\gamma_s}$, keeping the first order of $\Delta_c/\kappa$. Solving for $x$:
\begin{equation} \label{xlornew2}
 x=\frac{1}{\frac{2g}{C}\sqrt{\frac{\mu}{\gamma_s\gamma}}-1}\left(1-\frac{2i(\omega_c-\omega_L)}{\kappa}\right)+\frac{2i(\omega_L-\omega_0)}{\frac{\Delta_{\text{inh}}}{C}}.
\end{equation}
Using this, we can write our cavity reflection coefficient as 
\begin{align} \label{rco}
 r&=1-\frac{2\kappa_c}{(\kappa+2i(\omega_c-\omega_L))(1+x)} \nonumber \\
           &=1-\frac{\Delta_\text{CIT,min}\kappa_c/\kappa}{\Delta_\text{CIT}/2+i(\omega_L-\omega_\text{center})}
\end{align}

From this the center of CIT and width of CIT are:
\begin{equation}
    \omega_\text{center}=\left(\frac{\omega_0C}{\Delta_{\text{inh}}}-\frac{\omega_c}{\kappa}\right)\frac{\frac{\Delta_\text{inh}}{C}}{\left(1-\frac{\Delta_\text{inh}}{C\kappa}\right)}=\frac{1}{\left(1-\frac{\Delta_\text{inh}}{C\kappa}\right)}\omega_0-\frac{\frac{\Delta_\text{inh}}{C\kappa}}{\left(1-\frac{\Delta_\text{inh}}{C\kappa}\right)}\omega_c
\end{equation}
\begin{equation}
    \Delta_\text{CIT}=\frac{\frac{\Delta_\text{inh}}{C}}{\left(1-\frac{C}{2g}\sqrt{\frac{\gamma_s\gamma}{\mu}}\right)\left(1-\frac{\Delta_\text{inh}}{C\kappa}\right)}\rightarrow  \Delta_\text{CIT,min}=\frac{\frac{\Delta_\text{inh}}{C}}{\left(1-\frac{\Delta_\text{inh}}{C\kappa}\right)}
\end{equation}

The above derivations show the following:
\begin{enumerate}
    \item The CIT spectrum is similar to a cavity resonance, with $\Delta_\text{CIT}$ linewidth and input relative coupling ratio $\frac{\Delta_\text{CIT,min}\kappa_c}{\Delta_\text{CIT}\kappa}\rightarrow\frac{\kappa_c}{\kappa}$. This means that CIT is a mechanism in which a new resonance can be produced from a cavity, where this new resonance will be much narrower than the original cavity, whilst maintaining the input relative coupling ratio.
    \item This narrow feature can be used for precision sensing. The effective cavity (CIT) fluctuation is reduced by a factor of $\frac{\frac{\Delta_\text{inh}}{C\kappa}}{\left(1-\frac{\Delta_\text{inh}}{C\kappa}\right)}\rightarrow\frac{\Delta_\text{inh}}{C\kappa}$ compared to the original cavity fluctuation.
    \item The above derivation requires $\frac{\Delta_\text{inh}}{C\kappa}\ll1$ (besides the conditions in Eq.~\ref{AS}), which gives $\Delta_\text{inh} \ll 2\sqrt{N}g$. However, we can see that there is no individual requirement for $\kappa$, so the CIT phenomenon is not necessarily restricted to the bad cavity regime, and can work for cavities with higher quality factors. We note that if the cavity becomes extremely narrow, the mean-field approximation may not hold and a more exact model may be required. It is unclear, however, if we must be in the mean-field regime for CIT to occur.
    
\end{enumerate}
\begin{figure}[H]
    \centering
    \includegraphics[width=\linewidth]{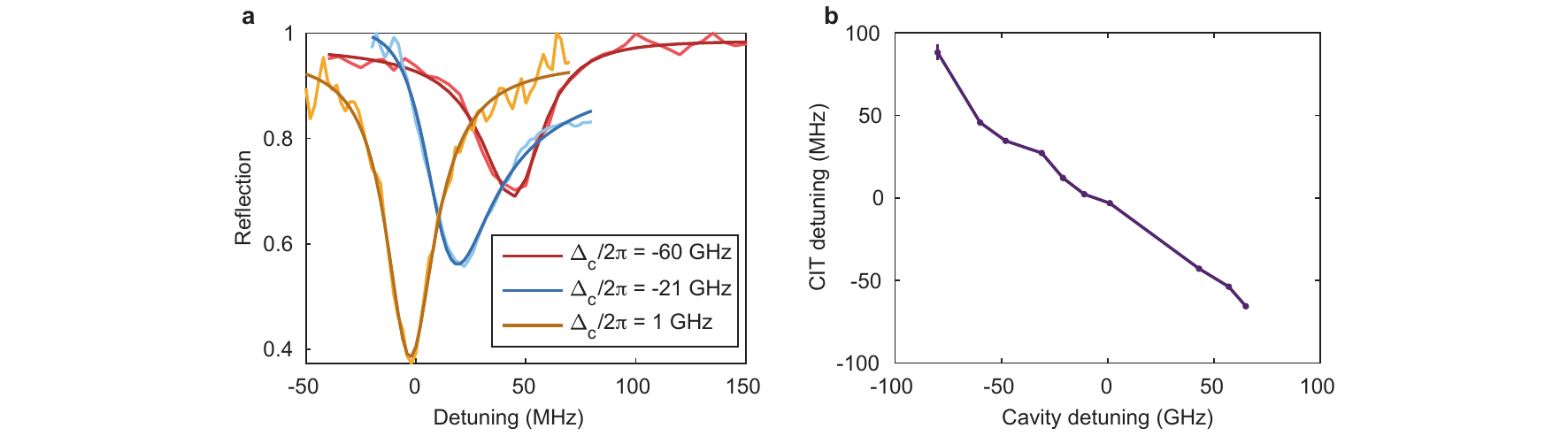}
    \caption{\textbf{a,} CIT spectrum for different cavity detunings, fit to a Fano resonance. \textbf{b,} Relation between CIT detunings and cavity detunings, showing the expected linear dependence. The error bar is from the fit.}
    \label{CIT_detuning}
\end{figure}

\subsubsection{Response time of CIT and application for optical switch}
In the previous sections, CIT has been investigated in the steady state regime. However, CIT has a finite response time, which is important to explore for certain applications such as an optical switch. To this end, we park the laser at the center of the CIT, and measure the reflection as a function of time (Extended Data FIG. 10).


We find a fast response time of $600$ ns, and posit that a two-port optical switch with an integrated filter can be realized using CIT. This current device has not been optimized for this particular application, hence we are unable to measure the other port (transmission) nor apply a transverse pump. Additionally, the contrast suffers from reflection due to an imperfect beam overlap with our optical coupler and our cavity being under-coupled. However, this problem can be solved by critically coupling and coupling the laser better into the cavity. This will result in increased contrast, important for improving the extinction ratio of the switch. For the best extinction ratio, the cavity should be designed as critical-coupled where $\kappa_1/\kappa=\kappa_2/\kappa=0.5$ where $\kappa_1$ and $\kappa_2$ are the coupling rate from port 1 and 2. We note that the pump light is to create the CIT, which is an absorptive process where the ions absorb light and get excited. In contrast, as long as the signal is weak enough, the optical switch can be non-absorptive. Furthermore, we believe the current switching time is limited by how fast the ions can reach saturation, which can be faster by engineering a larger $g$.

\subsubsection{Validity of the Tavis-Cummings model}
It has been shown in \cite{Blaha2020} that large collective coupling rates can lead to a breakdown of the standard Tavis-Cummings model. The relevant condition for the validity of the Tavis-Cummings model is $\text{FSR}\gg W$, where FSR is the free-spectral range of the cavity, and $W$ is the ensemble absorption derived in Eq.~\ref{Wabsorption}. Plugging in the parameters of our system, we find $\text{FSR}=2\pi\times25$ THz, and $W=2\pi\times0.5$ THz, thus satisfying the above inequality, and validating our use of the Tavis-Cummings model.

\section{Theoretical modelling for dynamics}
\subsection{Master equation formalism} \label{Master}
For simulating the dynamics of a driven inhomogeneous ensemble, we use the same Hamiltonian in Eq.~\ref{h}, and introduce various dissipative mechanisms through the Lindblad operators:
\begin{equation}
        \mathcal{L}_{\text{cav}}=\kappa(a\rho a^{\dagger} - \frac{1}{2}a^{\dagger}a\rho - \frac{1}{2}\rho a^{\dagger}a)
\end{equation}
\begin{equation}
        \mathcal{L}_{\text{em}}=\gamma_s\sum_{j=1}^N (\sigma_j^- \rho \sigma_j^+ - \frac{1}{2}\sigma_j^+ \sigma_j^- \rho - \frac{1}{2}\rho \sigma_j^+ \sigma_j^-)
\end{equation}
\begin{equation}
        \mathcal{L}_{\text{deph}}=\gamma_d\sum_{j=1}^N (\sigma_j^z \rho \sigma_j^z - \frac{1}{2}\sigma_j^z \sigma_j^z \rho - \frac{1}{2} \rho \sigma_j^z \sigma_j^z)=\gamma_d\sum_{j=1}^N (\sigma_j^z\rho \sigma_j^z - \rho)
\end{equation}
where $\mathcal{L}_{\text{cav}}$ is the cavity dissipation, $\mathcal{L}_{\text{em}}$ is the local spontaneous emission, $\mathcal{L}_{\text{deph}}$ is the local dephasing. $\gamma_s$ is the single atom spontaneous emission rate, and $\gamma_d$ is the excess dephasing rate.
We operate in the fast cavity limit, where $\kappa \gg g,\gamma_s,\gamma_d$, and adiabatically eliminate the cavity mode by setting $\dot{a}=0$. This allows us to replace the cavity field operator $a$ with atomic operators as:
\begin{equation} 
a=\frac{-i\sum_{j=1}^N g_j\sigma^-_j-\frac{\kappa}{2}\sqrt{\mu}}{i\Delta_c+\frac{\kappa}{2}}.
\end{equation}
We then rewrite $H$ and $\mathcal{L}_{cav}$ to $ H_{\text{at}}$ and $\mathcal{L}_{\text{col}}$ (collective dissipation) in terms of atomic operators:
\begin{equation}
     H_{\text{at}} = \frac{\Delta_c}{(\kappa/2)^2+\Delta_c^2} \sum_{j=1}^{N} g_j\sigma_j^+ \sum_{k=1}^{N} g_k\sigma_k^- + \Delta_c \mu + \frac{1}{2} \sum_{j=1}^{N} \Delta_j\sigma_j^z - \sum_{j=1}^{N}g_j\mu(\sigma_j^+ + \sigma_j^-)
\end{equation}
\begin{equation}
    \mathcal{L}_{\text{col}} = \frac{\kappa}{(\kappa/2)^2+\Delta_c^2}\sum_{j,k}^N g_jg_k(\sigma_j^- \rho \sigma_k^+ - \frac{1}{2}\sigma_j^+ \sigma_k^- \rho - \frac{1}{2}\rho \sigma_j^+ \sigma_k^-).
\end{equation}

We can further simplify $H_{\text{at}}$ and $\mathcal{L}_{\text{col}}$ by noting that $\kappa$ is large and the cavity is tuned to be on resonance with our atomic transition ($\Delta_c\rightarrow0$) as:

\begin{equation}
    H_{\text{at}} \approx \frac{1}{2} \sum_{j=1}^{N} \Delta_j\sigma_j^z - \sum_{j=1}^{N}g_j\mu(\sigma_j^+ + \sigma_j^-)
\end{equation}
\begin{equation}
    \mathcal{L}_{\text{col}} \approx \frac{4}{\kappa}\sum_{j,k}^N g_jg_k(\sigma_j^- \rho \sigma_k^+ - \frac{1}{2}\sigma_j^+ \sigma_k^- \rho - \frac{1}{2}\rho \sigma_j^+ \sigma_k^-).
\end{equation}
The system master equation now reads:
\begin{equation}
    \dot{\rho}=-i[H_{\text{at}},\rho]+\mathcal{L}_{\text{col}}+\mathcal{L}_{\text{em}}+\mathcal{L}_{\text{deph}}.
\end{equation}
We solve this system using QuTIP \cite{Johansson2013}. For simulations of system sizes greater than 8, we make use of Permutational Invariance (PIQS, \cite{Shammah2018}) enabling larger simulations. However, since this requires atoms to be identical, we stick to the full simulation in cases where we add inhomogeneities or want to look at correlations between ions.

\subsection{Excitation pulse length dependence}
\begin{figure}[H]
    \centering
    \includegraphics[width=0.65\linewidth]{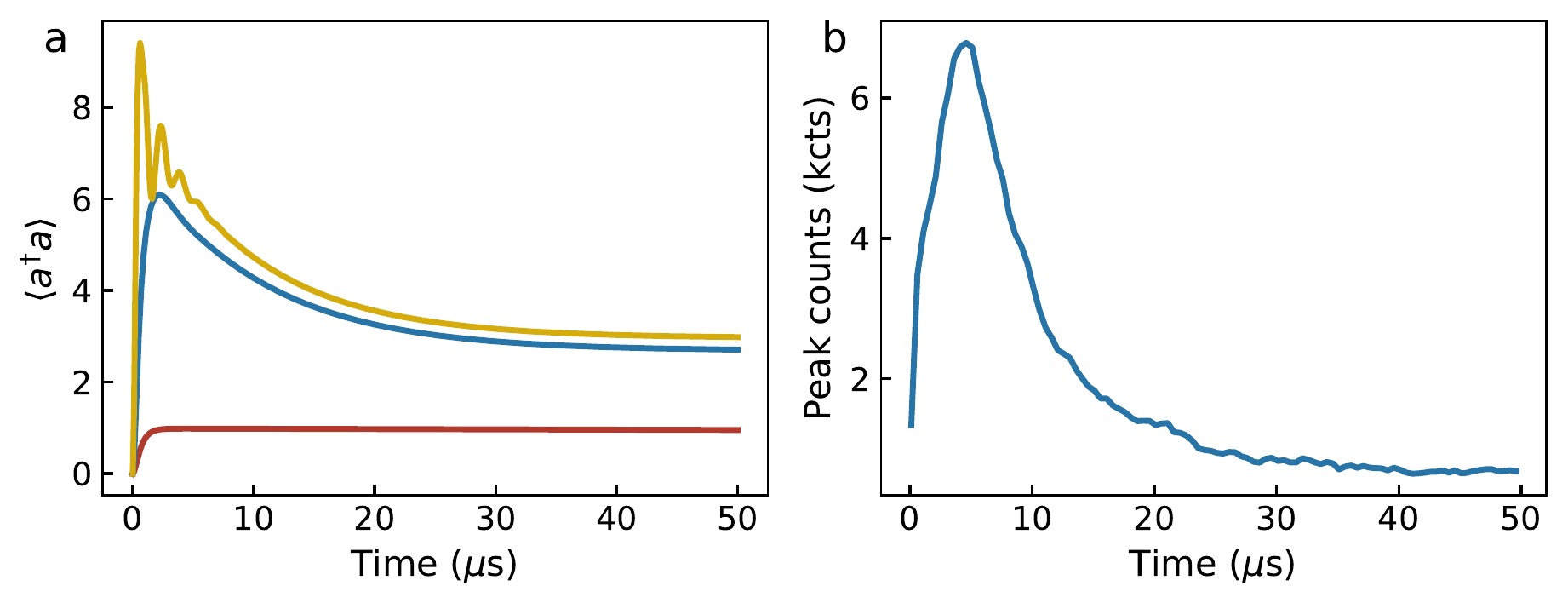}
    \caption{\textbf{a,} Simulation of the cavity field population $\langle a^{\dagger}a\rangle$ during a $50$ $\mu$s pulse applied to a system of $N=6$ identical ions, for different powers. The power increases from red (0.6 a.u.) to blue (1 a.u.) to yellow (1.5 a.u.) curves. Regardless of the power, here $50$ $\mu$s is long enough for the cavity to enter steady state. \textbf{b,} Experimental data for pulse length dependence of peak counts, with 10 nW of power, showing steady-state behavior at $50$ $\mu$s.}
    \label{plength}
\end{figure}

Here we study the non-steady state regime where the driving pulse length is short. We first theoretically investigate the atomic states during the driving period. In the case of $6$ identical atoms, we plot the normalized cavity mean photon number $\langle a^{\dagger}a\rangle\propto \langle J^+J^-\rangle$ during the drive for three different powers (Fig.~\ref{plength}a).

For a low drive (Fig~\ref{plength}a, red), we see that $\langle a^{\dagger}a\rangle$ quickly and monotonically increases to reach steady state. The steady-state population at this power consists primarily of lower excitation states, as the system is unable to be excited into the higher Dicke state manifolds. As we increase our drive, we see the appearance of Rabi-like oscillations among the Dicke states (Fig.~\ref{plength}a, yellow) and then a decay into steady state. The origin of this decay is two-fold: one is the dampening of the oscillations. This is due to collective decays (which go as $g^2/\kappa$) that drive the system towards equal population within a single vertical Dicke ladder. In other words, even in the absence of individual decay and dephasing, collective emission will spread the population out evenly among the superradiant states, leading to damped oscillations. Another decay type is the overall decay of the magnitude of $\langle a^{\dagger}a\rangle$. This is due to the coupling to the subradiant subspace through individual decay and dephasing, similar to the experimentally observed decrease in peak counts with power in regime II.

We explore this experimentally by measuring the excitation pulse length dependence of peak counts, shown in Fig.~\ref{plength}b. We find a qualitative match to the simulation of the initial build up of photons as we populate the superradiant states, and then a decay towards a steady state as we pump into the subradiant subspace. However, despite increasing the pump power, we do not observe the oscillations seen in Fig.~\ref{plength}a for high power (yellow). We attribute this to an averaging effect over our ensemble, as the oscillation frequency depends on $g$ and ion detuning, both inhomogeneous in our system.

\subsection{$N$-dependence of S-curve}
\begin{figure}
    \centering
    \includegraphics[width=1\linewidth]{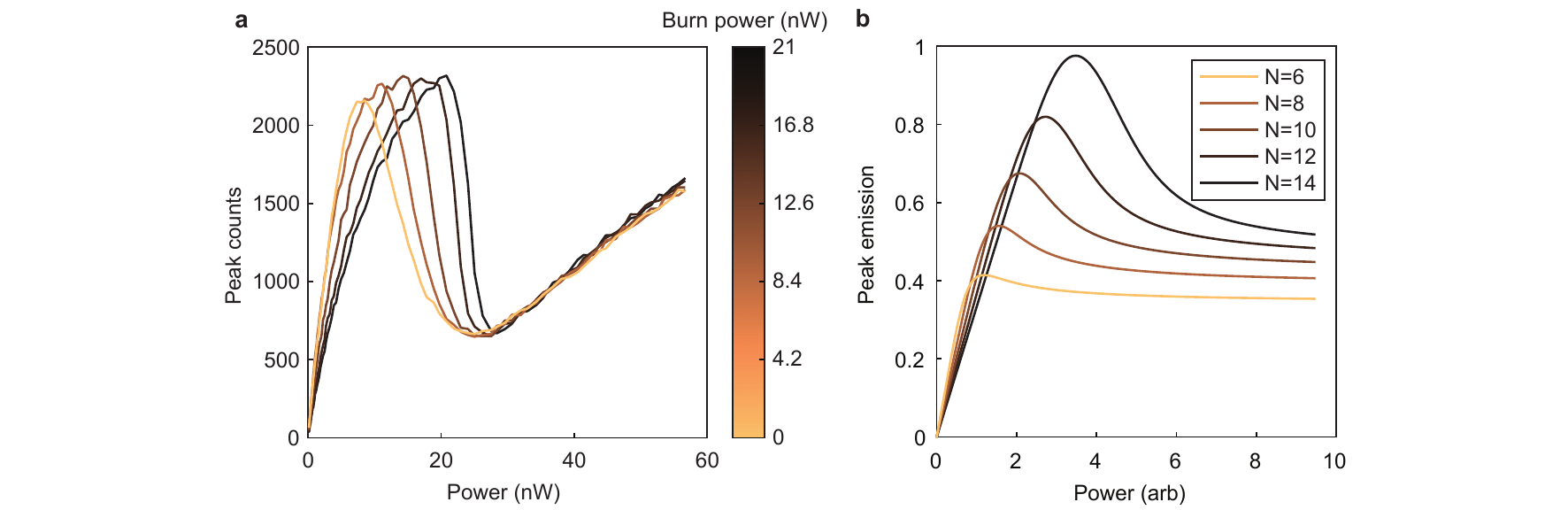}
    \caption{\textbf{a,} Extended data of Fig.~4b, showing more burning powers. For higher burning power, $N_\text{res}$ is larger and S-curve shifts towards the higher probe power. \textbf{b,} Simulation of varying ion number for the S-curve, showing a shift of the S-curve to higher powers. We find that, as expected, the turning point between regimes I and II shifts towards larger power, as our Dicke space expands and it requires more power to populate the subradiant subspace. As a result, at smaller powers, larger number of ions will have smaller peak emission, while the global maximum of peak emission increases for more ions.}
    \label{Ndep}
\end{figure}

Using PIQS, we simulate the dependence of the S-curve with varying system size $N$ (Fig.~\ref{Ndep}b). We find that as expected, the turning point between regimes I and II shifts towards larger power, as our Dicke space expands and it requires more power to populate the subradiant subspace. As a result, at smaller powers, larger number of ions will have smaller peak emission, while the global maximum of peak emission increases for more ions. This qualitatively matches what we see in the experiment (Fig.~\ref{Ndep}a). We note that regime III is not reproduced in this simulation as we do not consider the detuned ions.

\subsection{Comparison to the saturation effect}
An increase and subsequent decrease in emission can also occur due to the saturation of the atomic coherence (below, saturation effect), which occurs even for a single atom and is unrelated to the Dicke model. Here, we will show that we do not see the saturation effect, and explain the difference and the connection to the saturation effect. First, let us consider the single ion case, where we have already derived the analytical solution in Eq.~\ref{sigma-1} and Eq.~\ref{sigmaz1}.

\begin{figure}
    \centering
    \includegraphics[width=\linewidth]{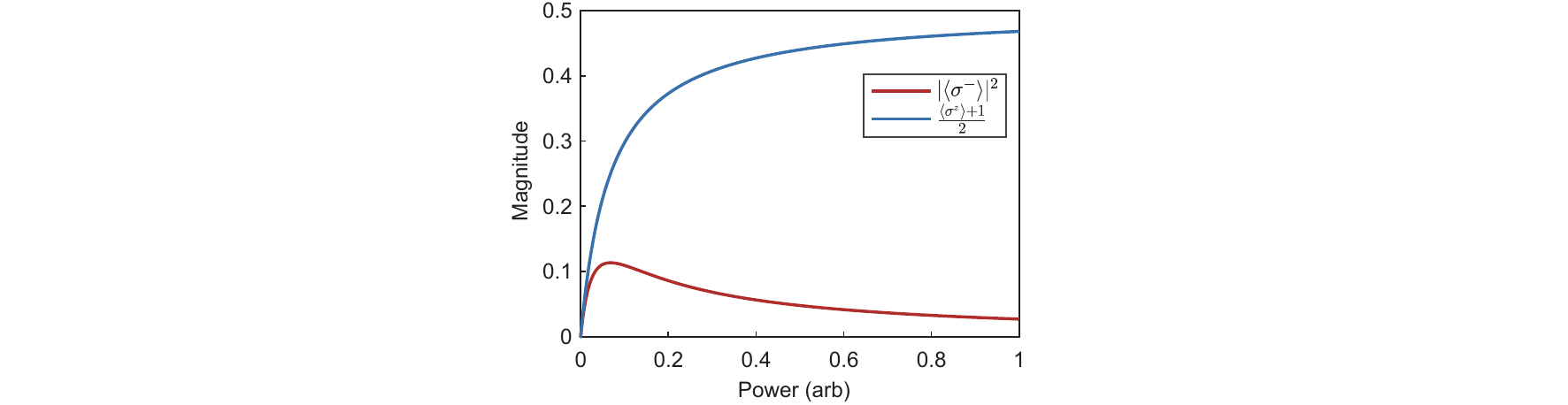}
    \caption{Comparison between excited state population, $|\frac{\langle\sigma_z\rangle+1}{2}|$, and the intensity of coherence, $|\langle\sigma^-\rangle|^2$, of a single atom as a function of excitation power.}
    \label{comsigmas}
\end{figure}

We can see that the intensity of the coherence ($|\sigma^-|^2$) first increases and then decreases with power where power $P \propto g^2\mu$. However, what is measured from cavity emission is 
\begin{equation}
    \langle a^+a^- \rangle =\frac{4g^2\langle\sigma^+\sigma^-\rangle}{\kappa^2}=\frac{4g^2}{\kappa^2}\frac{(\langle\sigma^z\rangle+1)}{2}
\end{equation}
Here we have applied Eq.~\ref{ae1} and the commutation rule. Then, by plotting both $|\sigma^-|^2$ and $|\frac{\sigma_z+1}{2}|$, we can see that while $|\sigma^-|^2$ has the non-monotonic trend, our measure $|\frac{\sigma_z+1}{2}|$ doesn't (Fig.~\ref{comsigmas}). In other words, since we are not directly measuring the coherence by detecting the cavity emission, our S-curve cannot be explained by saturation of coherence.

Once we move to the multiple ions case, the measure becomes
\begin{equation} \label{mulemit}
    \langle a^+a^- \rangle =\frac{4g^2J^+J^-}{\kappa^2}=
     \frac{4g^2}{\kappa^2}\left(
    \underbrace{\vphantom{\sum_{i\neq j}^N \langle \sigma_i^+ \sigma_j^- \rangle_\text{Correlation}}\sum_{i=1}^N \langle \sigma_i^+ \sigma_i^- \rangle}_\text{Individual} + \underbrace{\sum_{i\neq j}^N \langle \sigma_i^+ \sigma_j^- \rangle}_\text{Correlation}
     \right)
    =\frac{4g^2}{\kappa^2}\left(
     \underbrace{\vphantom{\sum_{i\neq j}^N \langle \sigma_i^+ \sigma_j^- \rangle_\text{Correlation}}\sum_{i=1}^N \left( \frac{\langle\sigma_i^z\rangle+1}{2} \right)}_\text{Individual} + \underbrace{\sum_{i\neq j}^N \langle \sigma_i^+ \sigma_j^- \rangle}_\text{Correlation} \right)
\end{equation}
as we analyzed in Eq.~17 in the main text. The first term is monotonically increasing relative to driving power, so the existence of the second term is necessary to observe S-curve shape. This second term is correlation between distinct ions, and will be absent for the single ion case, hence distinct from the saturation effects.

\section{Justification for photon emission fit function}
\begin{figure}[H]
    \centering
    \includegraphics[width=\linewidth]{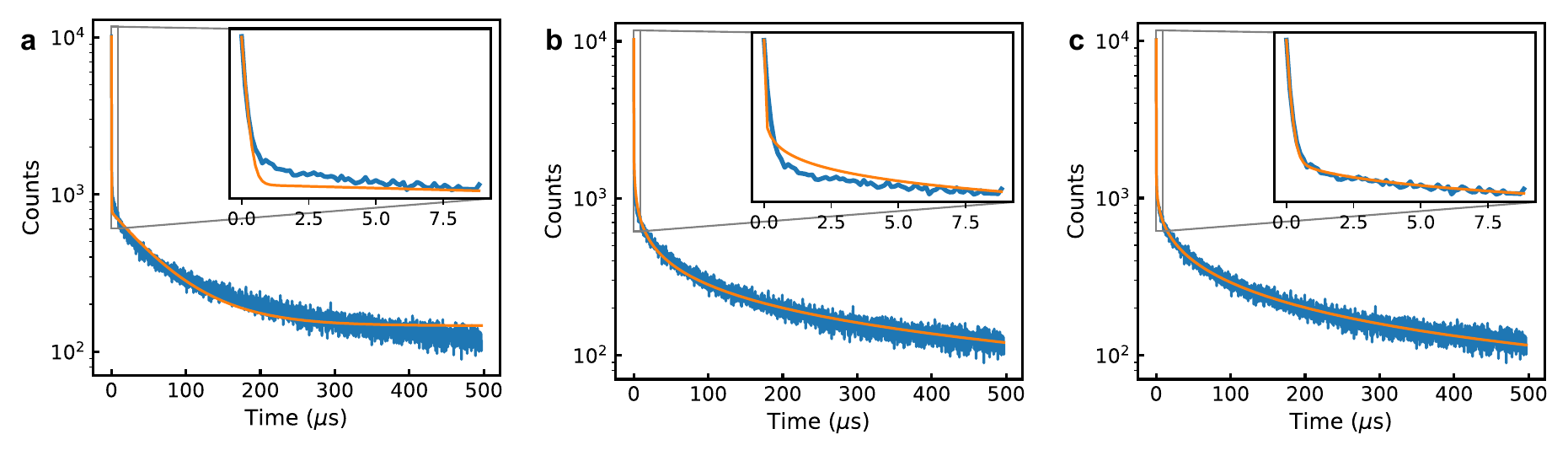}
    \caption{Fit of an emission curve to a \textbf{a,} Bi-exponential, \textbf{b,} Stretched exponential, and \textbf{c,} Addition of two stretched exponentials. The blue and orange lines are from experiments and fits, respectively. Insets show early-time comparisons between the data and the fits.}
    \label{fits}
\end{figure}

Here we provide justification for the function used in Fig.~3 to fit the decay curves. To find the best fit, one could start with a naive guess of a bi-exponential, to capture the superradiant and subradiant decays. A representative curve with a bi-exponential fit is shown in Fig.~\ref{fits}a. We see that the fit fails due to the strong multi-exponential nature of the subradiant decay. We then try a single stretched exponential fit in Fig.~\ref{fits}b. However, we still see that the fit fails at the crux between the slow and fast decay, shown in the inset.

To capture both the multi-exponential fast and slow decay, we now attempt a fast stretched exponential + slow stretched exponential (Fig.~\ref{fits}c). In particular, the fit function is $A_1\exp[-(t/\tau_1)^{x_1}]+A_2\exp[-(t/\tau_2)^{x_2}]+b$, where there is a fast decay with subscript 1 and slower decay with subscript 2. We find that this fits the data well in all power regimes, and most importantly does so with fit parameters that match our qualitative expectations from the picture provided in Fig.~3f in the main text. Additionally, while there are 7 fit parameters, we take several steps to minimize any overfitting of the data. First, the background $b$ is set to be the count level at a time much longer than the longest observed lifetime, essentially the dark counts and any leakage through our setup. Second, the existence of two decays is only relevant in regime II where there is a clear distinction between a fast and slow decay. In both regime I (III),  the fast (slow) decay component is dominant over the other, effectively resulting in just a single stretched exponential. 

\section{Estimation of the number of ions participating in superradiance} \label{sNions}
From Fig.~3d, we see that $x$ approaches 1 as the decay becomes single exponential for very low powers. In this case, we expect to be primarily exciting the single-excitation manifold, leading to a decay rate of $N\Gamma_c$, where $N$ is the number of atoms in our Dicke space, and $\Gamma_c$ is the Purcell enhanced decay rate of a single atom. While $N$ (and thus $\tau$) changes continuously in this regime due to power broadening, we still can estimate an effective Dicke space.

\begin{figure}
    \centering
    \includegraphics[width=0.35\linewidth]{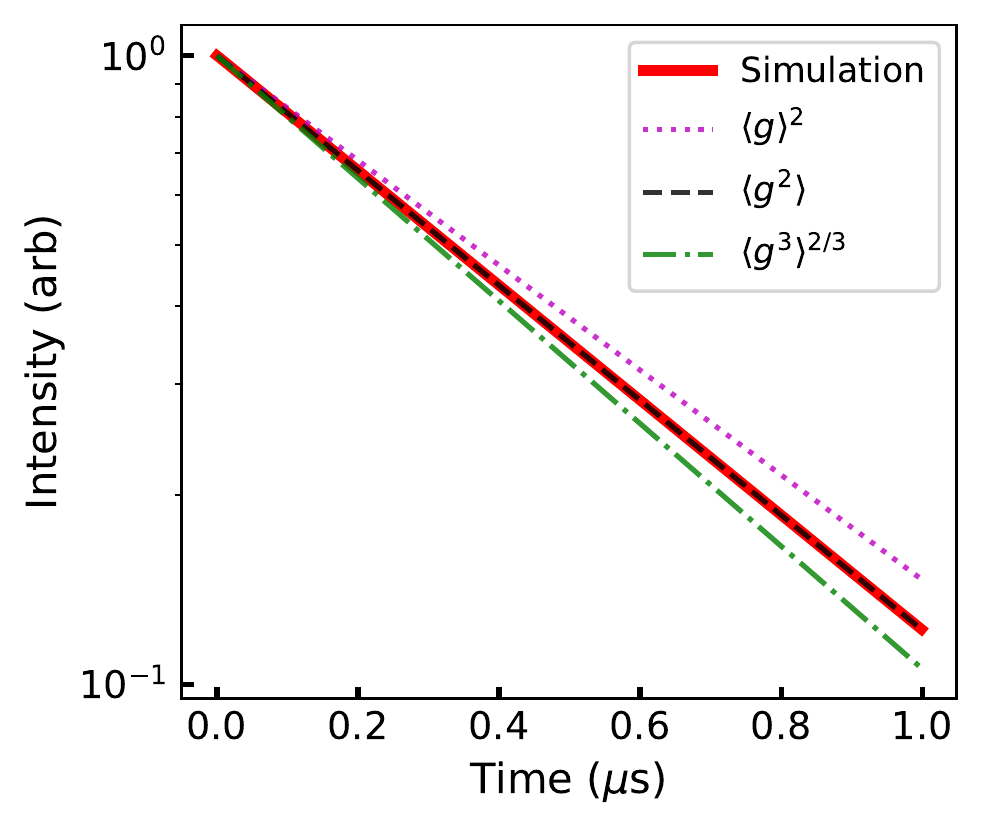}
    \caption{Master equation simulation of the decay of $N=6$ ions with inhomogeneous $g$ (red) with low excitation. The three dashed lines (pink, black, green) are single exponential decays with decay rate $\frac{4N\langle g\rangle^2}{\kappa}$, $\frac{4N\langle g^2\rangle}{\kappa}$, $\frac{4N\langle g^3\rangle^{2/3}}{\kappa}$, respectively. The results show that, as expected, the superradiant decay from the single excitation manifold is governed by the average of $g^2$ for the case of inhomogeneous $g$.}
    \label{ginhomo}
\end{figure}

In order to estimate $N$, we first show that the superradiant decay rate is $N$ times the average Purcell decay rate, given that our system has inhomogeneous $g$. To this end, we simulate the time dynamics of $6$ ions with varying $g$, excited with low power (Fig.~\ref{ginhomo}). We find that as expected, the superradiant decay time from the lowest manifold is given by $\frac{4N\langle g^2\rangle}{\kappa}$, indicated by the overlap of the simulation and analytical time decay curves. From Section 1, we estimate the average Purcell enhancement by using $\sqrt{\langle g^2 \rangle}=2\pi\times10.6$ MHz, giving an average Purcell decay time of $15.6$ $\mu$s. Meanwhile, we measured a decay time of $270$ ns in the low power regime with $x\approx1$ (Fig.~3d), from which we can estimate the effective number of ions participating in superradiance at this particular power to be $\sim50$.

\section{Cooperativity required to observe superradiance}
For a general symmetric distribution of ions, $\frac{iW(\omega=\omega_0)}{N\rho(\omega=\omega_0)}=\pi g^2$ \cite{Diniz2011} where $\rho(\omega=\omega_0)$ is the ion probability distribution at $\omega=\omega_0$, from which we obtain $C=\frac{|W(\omega=\omega_0)|}{\kappa/2}=\frac{2\pi g^2 N\rho(\omega=\omega_0)}{\kappa}$.
This ensemble cooperativity represents the ratio of the absorption rate to the cavity decay rate, which indicates the number of ions a photon can interact with before it leaks out of the cavity. The condition to observe superradiance in an inhomogeneous ensemble is roughly given by $\tau_R<T_2^*$ \cite{Temnov2005}, where $\tau_R$ is the slowest superradiant decay time in the absence of inhomogeneity, and $T_2^*$ is the inhomogeneous dephasing time. Assuming there are $N_{\text{eff}}$ ions participating in superradiance, spanning a frequency $\Delta_{\text{eff}}$, this leads to the requirement that the superradiance decay rate $N_{\text{eff}}\frac{4g^2}{\kappa}$ must be larger than the effective bandwidth $\Delta_{\text{eff}}$ of the participating ions. For the ions around $\omega_0$, we know $N_{\text{eff}}=\Delta_{\text{eff}}N\rho(\omega=\omega_0)$. Using $N_{\text{eff}}\frac{4g^2}{\kappa}>\Delta_{\text{eff}}$, we obtain $\frac{4g^2N\rho(\omega=\omega_0)}{\kappa}>1$, giving an estimate of the cooperativity required to observe superradiance as $C>\frac{\pi}{2}$.


\section{Cavity emission and reflection of the entire spectrum}
\begin{figure}
    \centering
    \includegraphics[width=\linewidth]{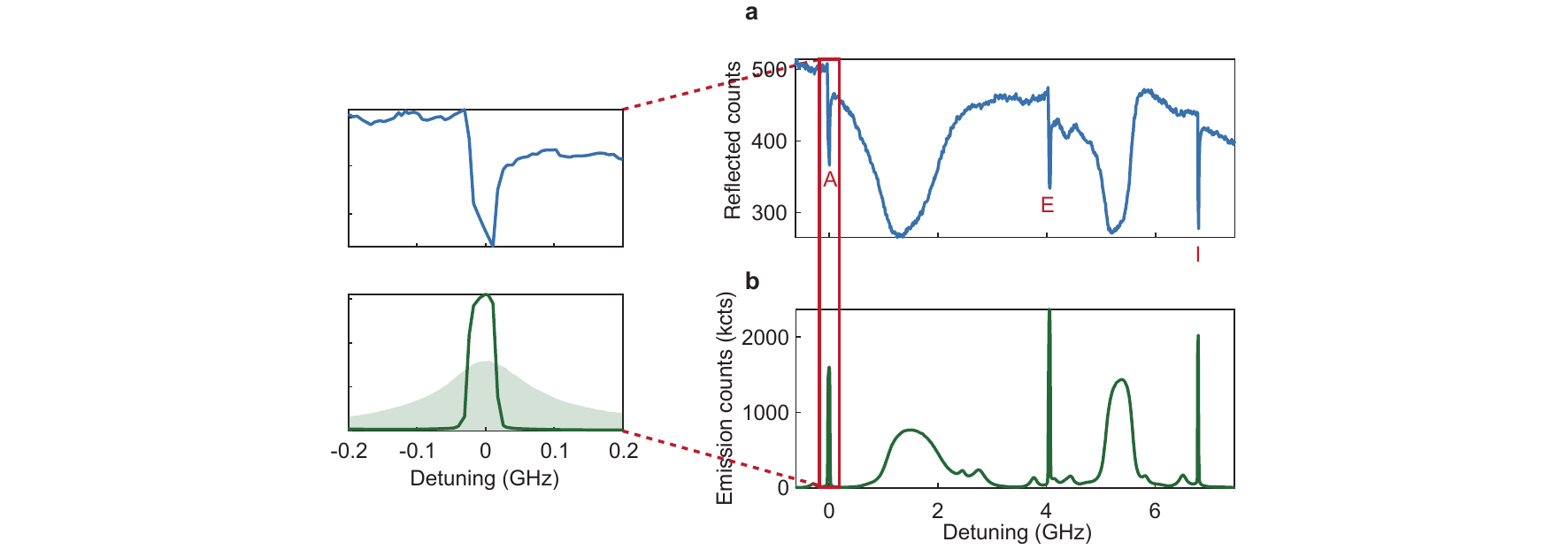}
    \caption{Pulsed excitation with $47$ nW as a function of laser detuning, where the \textbf{a,} reflected pulse amplitude and \textbf{b,} integrated emission counts after the pulse are extracted. The x-axis denotes the laser detuning relative to A transition. The other sharp peaks at $\sim4$ and $\sim7$ GHz correspond to E and I transitions, respectively. The green shaded area indicated the inhomogeneous ion distribution, clearly showing that the emission is nonlinear and narrower than the ion distribution due to phase cancellation.}
    \label{PLall}
\end{figure}

Here we show pulsed excitation measurements at different laser frequencies, covering all of the transitions (Fig.~\ref{PLall}). The laser frequency is swept from the lower frequency side of the A transition to the higher frequency side of I. In Fig.~\ref{PLall}a we plot the reflected pulse amplitude and in Fig.~\ref{PLall}b we show the integrated total emission counts. There are two aspects of the plot in Fig.~\ref{PLall}b. Firstly, we find three tall peaks corresponding to A, E, and I. Zooming into one of these transitions, we can see that the spectral width of the emission is narrower than the inhomogeneous linewidth of the ensemble, and resembles more a flat top rather than a Gaussian/Lorentzian (see zoom in of Fig.~\ref{PLall}b). This is because CIT creates a transparency window in the center of the inhomogeneous line, allowing more light to enter through that window. Thus, the CIT shapes the incoming light with width approximately equal to the CIT width (zoom in for Fig.~\ref{PLall}a). Secondly, between the A and E, and E and I transitions, at detuning around $1.5$ and $5.5$ GHz, there are elevated counts where there are no resonant ions. This is related to the phase cancellation effect observed in CIT. Specifically, when there are balanced ions on either side of the laser frequency, more light can enter the cavity, resulting in enhanced off-resonant excitation for the same input power.

\bibliography{ref_supplementary}
\bibliographystyle{naturemag}